\documentclass[bibyear]{aa}

\usepackage{natbib}
\usepackage{graphicx}
\usepackage{txfonts}
\usepackage{color}
\usepackage{longtable}
\usepackage{pdflscape}
\usepackage{rotating}
\usepackage{ amssymb }

\newcommand{\msun}{\mbox{$\,{\rm M}_\odot$}}

\begin{document}

\title{Uncrowding R\,136 from VLT/SPHERE extreme adaptive optics \thanks{Based on data collected at the European Southern Observatory, Chile, Guaranteed Time Observation 095.D-0309(K)}}
\author{ 
        Z. Khorrami\inst{1,}\inst{2}
        \and
        F. Vakili \inst{1}
        \and
        T. Lanz \inst{1}
        \and
        M. Langlois \inst{3,}\inst{4}
        \and
        E. Lagadec \inst{1}
        \and
        M. R. Meyer \inst{5,} \inst{6}
        \and
        S. Robbe-Dubois \inst{1} 
      \and
        L. Abe \inst{1}
       \and
        H. Avenhaus \inst{7}
       \and
        JL. Beuzit  \inst{8}
       \and
        R. Gratton \inst{9}
       \and
                D. Mouillet \inst{8}
        \and
        A. Orign\'e  \inst{4}
       \and
       C. Petit \inst{10}
        \and
        J. Ramos \inst{11}
}

\institute{
 Universit\`e C\^ote d'Azur, OCA, CNRS, Lagrange, France
 \email{KhorramiZ@cardiff.ac.uk}
 \and
 School of Physics and Astronomy, Cardiff University, The Parade, Cardiff CF24 3AA, UK
        \and
        Univ Lyon, Univ Lyon1, Ens de Lyon, CNRS, CRAL UMR5574, F-69230, Saint-Genis-Laval, France
        \and
        Aix Marseille Université, CNRS, LAM - Laboratoire d’Astrophysique de Marseille, UMR 7326, 13388, Marseille, France
        \and
        Institute for Astronomy, ETH Zurich, Wolfgang-Pauli-Strasse 27, CH-8093 Zurich, Switzerland
        \and
        Department of Astronomy, University of Michigan, Ann Arbor, MI 48109, U.S.A.
        \and
        Departamento de Astronomìa, Universidad de Chile, Casilla 36-D, Santiago, Chile
      \and
      Université Grenoble Alpes, CNRS, IPAG, 38000 Grenoble, France
       \and
       INAF - Astronomical Observatory of Padua, Italy	
       \and
       ONERA - Optics Department, 29 avenue de la Division Leclerc, F-92322 Chatillon Cedex, France
\and
Max-Planck-Institut fur Astronomie, Konigstuhl 17, D-69117 Heidelberg, Germany
}

\date{}

\abstract{
\textit{Context.} This paper presents the sharpest near-IR images of the massive cluster R\,136 to date,
based on the extreme adaptive optics of the SPHERE focal instrument implemented on the ESO Very Large Telescope and operated in its IRDIS  imaging mode.
 
 \textit{Aims.} The crowded stellar population in the core of the R\,136 starburst compact cluster remains still to be characterized in terms of individual luminosities, age, mass and multiplicity. SPHERE/VLT and its high contrast imaging possibilities open new windows to make progress on these questions.
 
\textit{Methods.} Stacking-up a few hundreds of short exposures in J and Ks spectral bands over a Field of View (FoV) of 10.9"$\times$12.3" centered on the R136a1 stellar component, enabled us to carry a refined photometric analysis of the core of R\,136.  We detected 1110 and 1059 sources in J and Ks images respectively with 818 common sources.

\textit{Results.} Thanks to better angular resolution and dynamic range, we found that more than 62.6\% (16.5\%) of the stars, detected both in J and Ks data, have neighbours closer than 0.2" (0.1").

The closest stars are resolved down to the full width at half maximum (FWHM) of  the point spread function (PSF) measured by {\it{Starfinder}}.
Among resolved and/or detected sources R136a1 and R136c have optical companions and R136a3 is resolved as two stars (PSF fitting) separated by $59\pm2$ mas. This new companion of R136a3 presents a correlation coefficient of 86\% in J and 75\% in Ks. 
The new set of detected sources were used to re-assess the age and extinction of R\,136 based on 54 spectroscopically stars that have been recently studied with HST slit-spectroscopy (Crowther et al. 2016) of the core of this cluster. 

Over 90\% of these 54 sources identified visual companions (closer than 0.2").
We found the most probable age and extinction for these sources are $1.8^{+1.2}_{-0.8}$ Myr, $A_J=(0.45\pm0.5)~mag$ and $A_K=(0.2\pm0.5)~mag$ within the photometric and spectroscopic error-bars.
Additionally, using PARSEC evolutionary isochrones and tracks, we estimated the stellar mass range for each detected source (common in J and K data) and plotted the generalized histogram of mass (MF with error-bars). 

Using SPHERE data, we have gone one step further and partially resolved and studied the IMF covering mass range of (3 - 300) \msun~ at the age of 1 and 1.5 Myr. The density in the core of R\,136 (0.1 - 1.4 pc) is estimated and extrapolated in 3D and larger radii (up to 6pc).
We show that the stars in the core are still unresolved due to crowding, and the results we obtained are upper limits. Higher angular resolution is mandatory to overcome these difficulties.   
}
   \maketitle
\section{Introduction}
{\textit{"The two Magellanic clouds, Nubecula major and Nebecula minor, are very remarkable objects... 
In no other portion of the heavens are so many nebulous and stellar masses thronged together in an equally small space."}} \footnote{From a letter of Sir John Herschel, to Feldhuysen, at the Cape of Good Hope, 13 June, 1836.}

R\,136 is a very massive young star cluster that lies at the center of the Tarantula nebula in the  Large Magellanic Cloud (LMC). Hosting the most massive stars known in the Local Universe (Crowther et al. 2016, Crowther et al. 2010), R\,136 provides a unique opportunity to study the formation of massive stars and clusters in the early stages of evolution. 

Our understanding of the true nature of R\,136 has constantly improved with increasing telescope resolution. The fuzzy object in the core of the cluster was initially thought to be a super-massive star with a mass in excess of 1000 M$_{\odot}$ (Feitzinger et al. 1980; Cassinelli et al. 1981; Savage et al. 1983). Image-sharpening techniques such as speckle interferometry (Weigelt \& Baier 1985) revealed  that this object had many individual stars, however, which ere coined the R136a Weigelt components. R136a was clearly resolved into hundreds of sources by the Hubble Space Telescope (HST, Hunter et al. 1995; Campbell et al. 1992), which opened routes to better study each of Weigelt components at various wavelengths and resolutions. More recent multiconjugate adaptive optics instruments (AO, Campbell et al. 2010) on the VLT attempted to better resolve the core of R\,136 with relative success. 

The combination of photometry, ultraviolet spectrometry (STIS/MAMA, Crowther et al. 2016), visible (HST/FOS, Massey \& Hunter 1998) and near-infrared (VLT/SINFONI, Schnurr et al. 2009) observations has resulted in more constraints on the R136 stellar population and its most luminous stars. 
Still, questions persist on the true nature of these stars: their multiplicity, their mass and the age of the cluster as a whole are still unclear.  

This paper presents the first observations of R\,136 in the infrared by the second-generation ESO Very Large Telescope (VLT) instrument Spectro-Polarimetric High-contrast Exoplanet Research (SPHERE \footnote{https://www.eso.org/sci/facilities/paranal/instruments/sphere.html}) (Beuzit et al. 2008) aiming at uncrowding the dense central region of R\,136 in the near-infrared (near-IR). 
Thanks to SPHERE's extreme AO system, we reached the same resolution as HST in the visible with the VLT 8.2m Melipal telescope in the K$_s$ band and have surpassed it in J band thanks to the angular resolution, namely  0.035-0.055 arcsec in near-IR and with a better pixel sampling (12.25 mas/pix).

\section{Observations}
\begin{figure*}
\centering
\includegraphics[trim=0 10 700 60,clip,width=8.5cm]{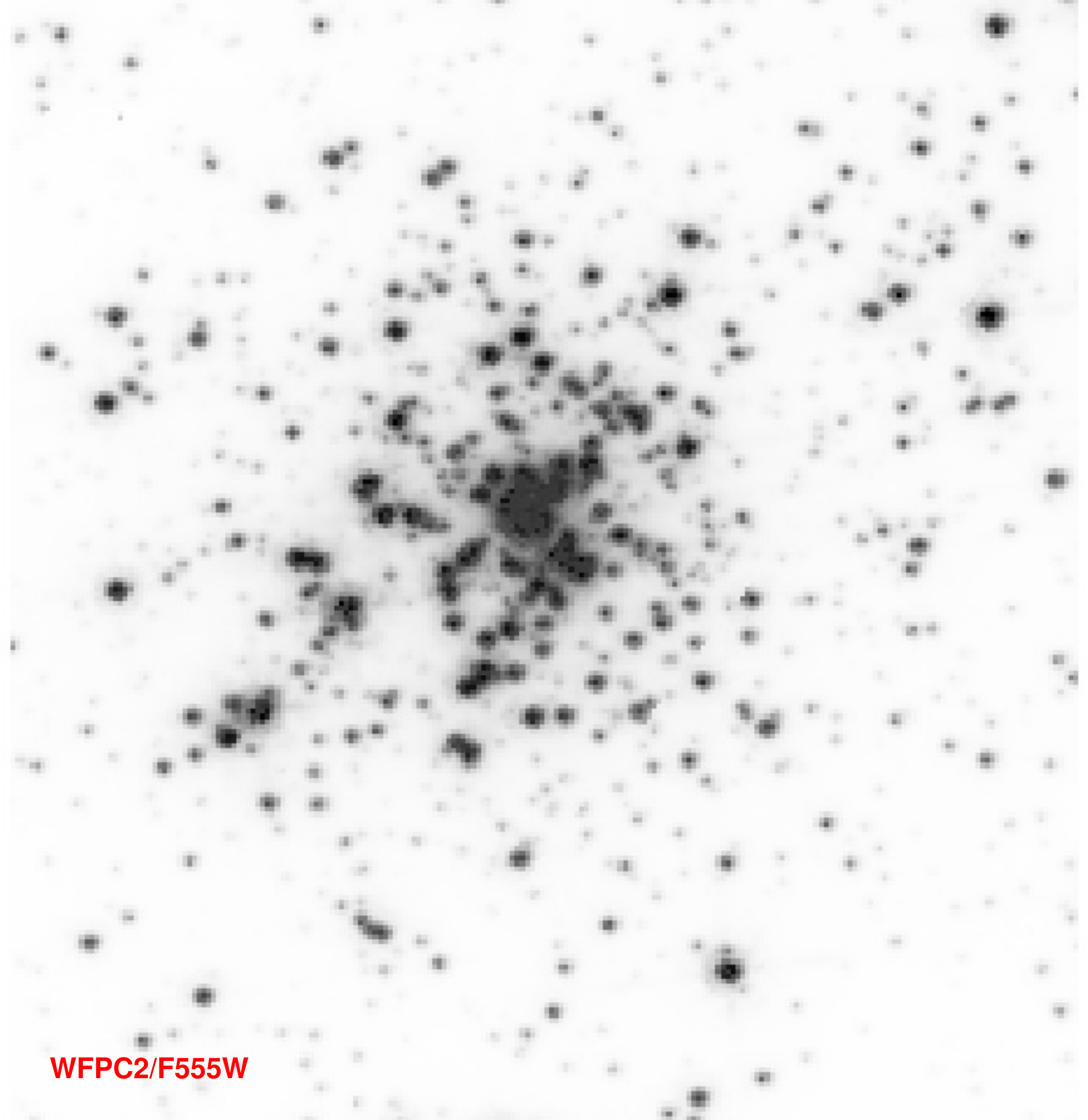}
\includegraphics[trim=0 0 0 0,clip,width=8.5cm]{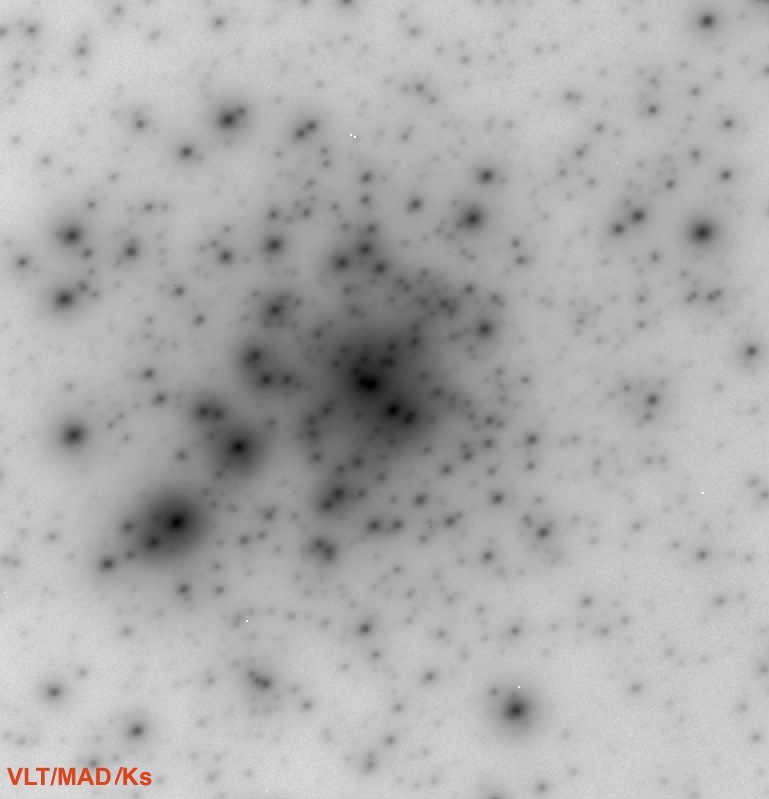}\\
\includegraphics[trim=0 0 0 0,clip,width=8.5cm]{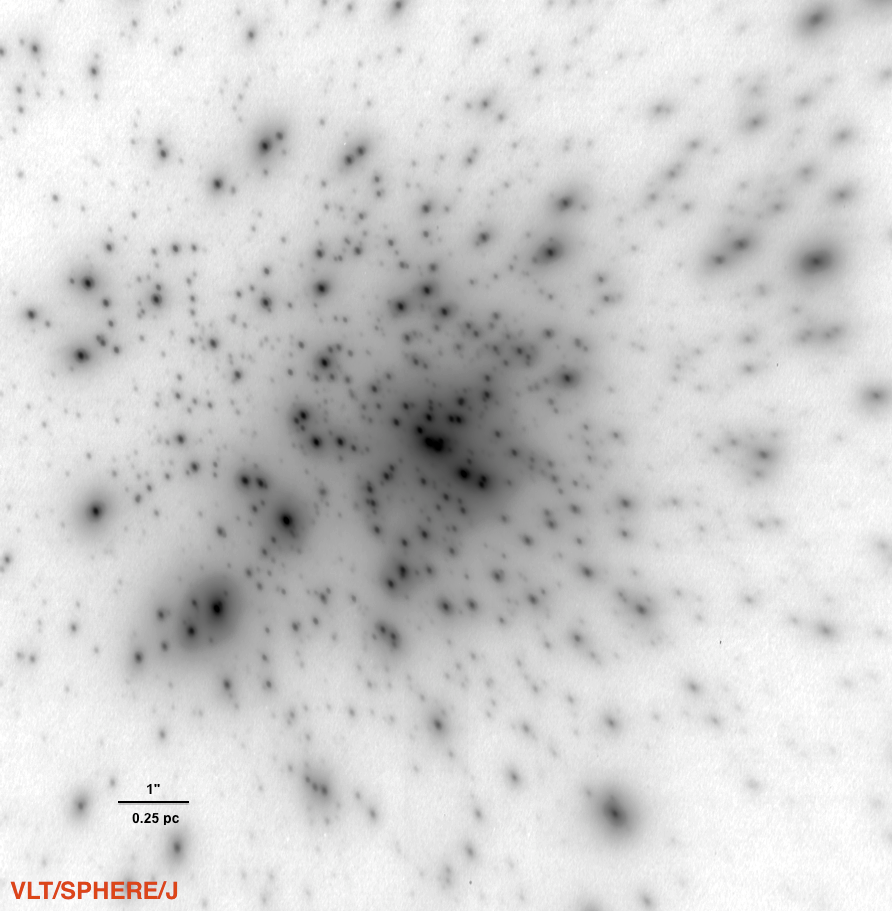}
\includegraphics[trim=0 0 0 0,clip,width=8.5cm]{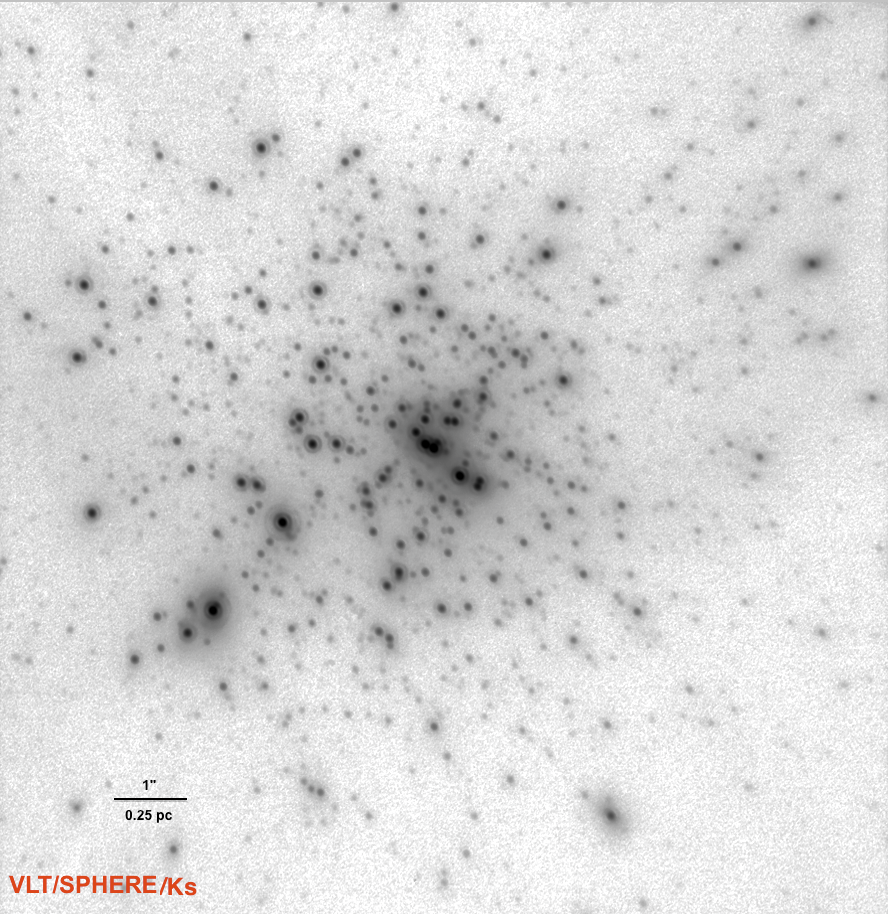}
\caption{Comparison of R136 core images at different wavelengths with the highest available angular resolution telescopes. The FoV of all images is the same as IRDIS data (10.9"$\times$12.3"). Top left: HST/WFPC2 in V band ($\lambda_{cen}$526nm), Top right: VLT/MAD in Ks band ($\lambda_{cen}$2200nm), bottomn right: SPHERE/IRDIS/Ks ($\lambda_{cen}$2182nm); bottom left: SPHERE/IRDIS/J ($\lambda_{cen}$1245nm)}
\label{fig:hst555}
\end{figure*}

We collected data in Guaranteed Time Observation (GTO) runs to image R\,136 using the classical imaging mode of IRDIS (Langlois et al. 2014).
For our purpose we use the same spectral band split into two channels to correct for residual detector hot pixels and uncorrelated detector noise among other instrumental effects. 
With this method we maximized the image sharpness of the total exposure by discarding the single frames with poorer Strehl ratio and corrected the residual tip-tilt error on each short exposure through software.
Observations were performed in September 2015, with high dynamic and high angular resolution imaging in J and Ks bands, over a field of view (FoV) of 10.9"$\times$12.3", centered on the core of the cluster (Figure \ref{fig:hst555} bottom). The seeing was $0.63 \pm 0.1"$ during the observations. The night was rated as clear. Less than 10\% of the sky (above 30 degree elevation) were covered in clouds, and the transparency variations were lower than 10\%.

In order to qualify our data, we compared the reduced J- and Ks-band images with published images of R\,136 from HST (WFPC2 and WFC3) in V band and the VLT/MAD imaging (Campbell et al. 2010) in K band (Figure \ref{fig:hst555} top) . 
This comparison confirms that our data present better spatial resolution and point-spread function (PSF) sampling more suitable for applying deconvolution techniques.

Our data consist of 300 frames of 4.0s exposures in the two IRDIS broad-band Ks and J filters (BB-Ks, BB-J).
The Wolf-Rayet star R136a1 was used to guide the AO loop of SPHERE, confirming its better-than-nominal performances that surpass NACO and MAD observations.

The range of airmass during these observations was 1.54 to 1.67. A log of observations is presented in Table \ref{table:expo}. 
\begin{table}[h!]
\caption{Exposure time log of VLT/SPHERE observations of R136.}
\centering
\begin{tabular}{ c c c c c  c}  
Obs. date& Filter&Single/Total &$\lambda_{cen}$[nm] &$\Delta\lambda$ [nm] \\
    & & Exposure[s] && \\
\hline
2015-09-22 &BB-J &4.0/1200 &1245&240\\
2015-09-22 &BB-K&4.0/1200 &2182&300\\
\end{tabular} 
\label{table:expo}  
\end{table}
We used the SPHERE pipeline package \footnote{\url{http://www.mpia.de/SPHERE/sphere-web/nightly_builds-page.html}} to correct for dark (current), flat (fielding), (spatial) distortion, bad pixels and thermal background due to instrument and sky (in Ks). 
Furthermore and in order to reach the highest sensitivity and the largest number of detectable sources, additional corrections were carried out on the images. Based on a Gaussian fit using selected stars, we estimated and corrected the subpixel image drifts before combining the individual images. 
This allowed us to correct for residual tip-tilt errors with an accuracy of a few mass. Still, some uncorrected atmospheric leaks persist in our final images; they are due to the adaptive optics residual halo, which is stronger in J than in Ks. We accurately considered this to estimate correct error bars in addition to the {\it{Starfinder}} reduction tool, which provided photometric errors (see Section \ref{sec:photometry}).

\section{Photometry of the R136 compact core}\label{sec:photometry}

We used the {\it{Starfinder}} package implemented in IDL (Diolaiti et al. 2000). {\it{Starfinder}} is designed for the analysis of AO images of crowded fields such as the Galactic Center (Pugliese et al. 2002). 
It determines the empirical local PSF from several isolated sources in the image and uses this PSF to extract other stellar sources across the FoV.
{\it{Starfinder}} also estimates the formal error on the estimated photometry based on photon noise, variance due to the sky, and the PSF fitting procedure itself. This error is called {\bf{"PSF fitting error"}} hereafter.

Our photometry analysis of R\,136 was conducted in two steps: 1) stellar source detection using {\it{Starfinder}} and 2) the background analysis to obtain realistic error bars on the photometry of individual stars beyond the formal PSF fitting error.

\begin{figure}
\centering
\includegraphics[trim=30 0 20 0,clip,width=8.cm]{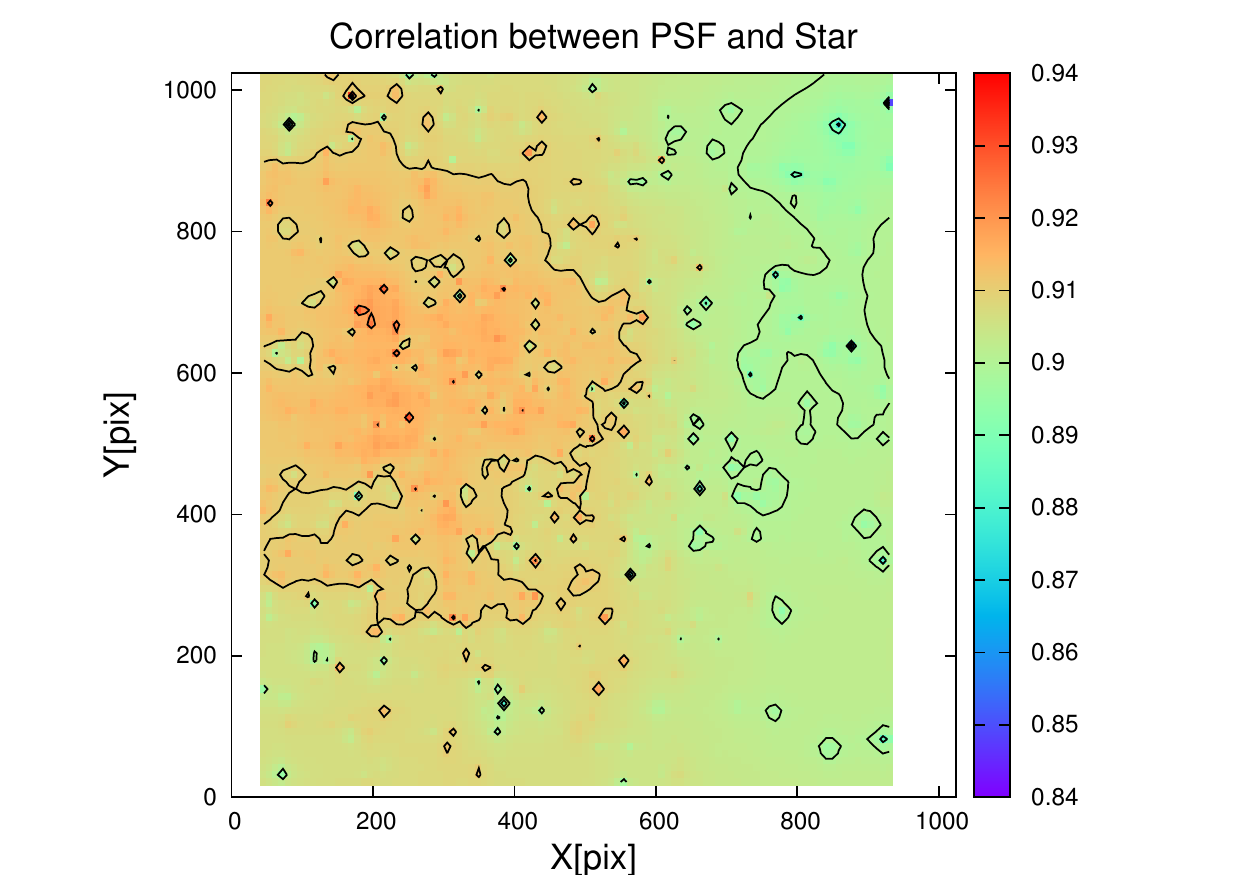}\\
\includegraphics[trim=30 0 20 18,clip,width=8.cm]{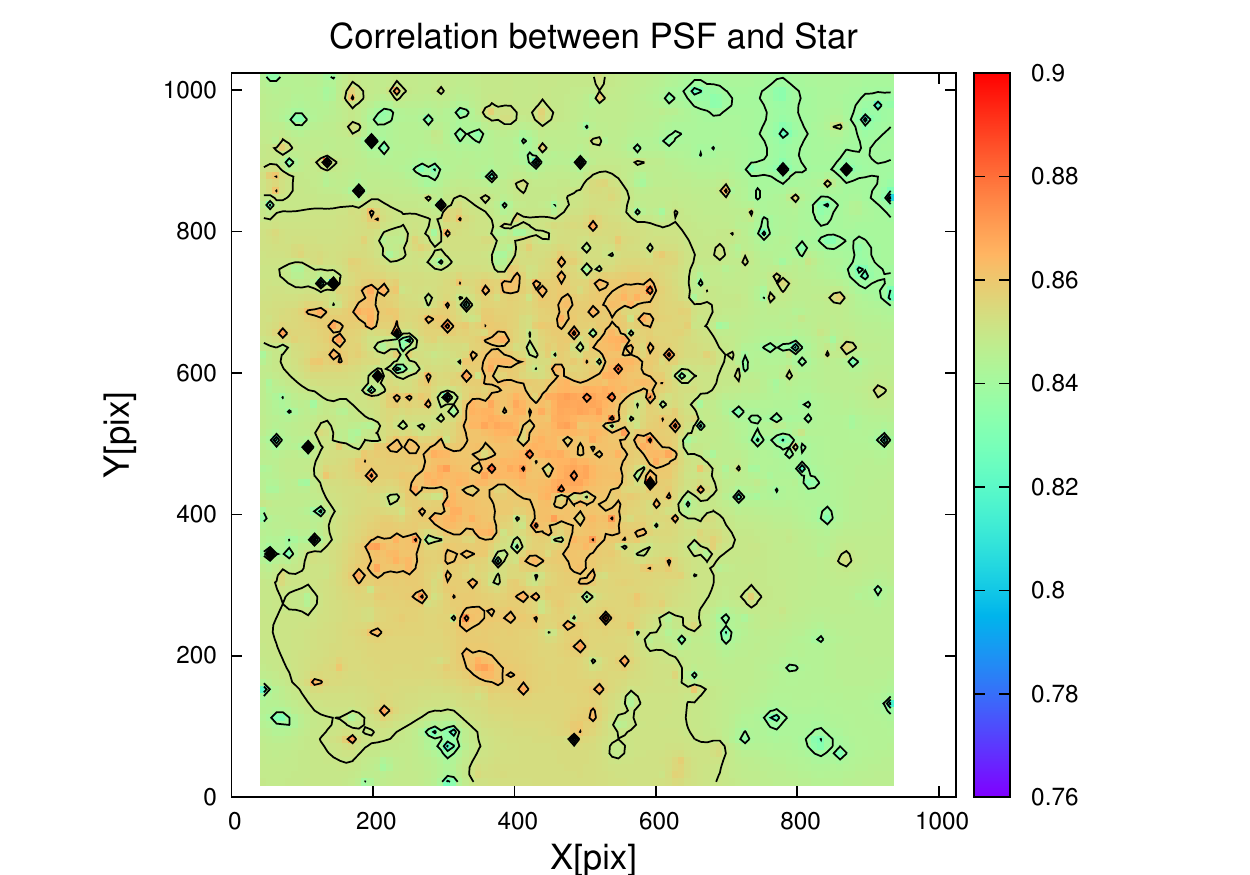}
\caption{Map of the correlation coefficients between the input PSF and each detected stellar source, calculated by {\it{Starfinder}} across the FoV in J (top) and in K (bottom) images of R\,136 by SPHERE.}
\label{fig:mapcorre}
\end{figure}

In the first step, three well-isolated (within $0.47"\times0.47"$) stars were used to extract the PSF in J- and Ks-band data separately. The extracted PSFs from J- and Ks-band data were used as input for stellar source detection by {\it{Starfinder}}. The FWHM of the extracted PSFs are 54.71 and 65.16 mas in J- and Ks-band data, respectively. Consequently, 1110 and 1059 sources were detected in the J and Ks bands, respectively. We stopped source extraction after we attained a minimum correlation coefficient of 65\% and 80\% in Ks and J bands between the extracted star with the locally determined PSF according to {\it{Starfinder}} procedures. Stars with higher correlation coefficients, that is, with more similarity to the PSF, represent a higher reliability of their photometric measure. 
Figure \ref{fig:mapcorre} shows the map of the correlation coefficient across the field of J (left) and Ks (right) images of R136. The PSF changes as a function of radial distance and azimuth along the field  (Figure \ref{fig:mapcorre}).
The AO correcting efficiency degrades as a function of distance from R136a1, which is the reference star for the AO loop. At the borders of the FoV, the isoplanatic limits are approached.
Overall, the PSF is not centro-symmetric at large distances from R136a1. We took these distortions into account to estimate the local statistical errors, which become significant on the distant sources from the center of the image, typically >3".

In addition to the correlation coefficient criterion, we applied the limit of standard deviation from the sky brightness ($\sigma_{sky}$) to stop the extraction of sources by {\it{Starfinder}}, that is, the local PSF maximum value must exceed $2\sigma_{sky}$ over the sky. 
The faintest common detected stars between J and Ks band have a signal-to-noise ratio (S/N) better than 2.

The Strehl ratio (SR) in J and Ks is determined as $0.40 \pm 0.05$ and $0.75 \pm 0.03$, respectively, on average. These estimates are assessed from the SPARTA files recorded during simultaneous SPHERE runs with AO-corrected images of R136. They are based on the slope measurements of the Shack-Hartmann wavefront sensor in SAXO by extrapolating the phase variance deduced from the reconstruction of SAXO open-loop data using deformable mirror, tip-tilt voltages, and wavefront sensor closed-loop measures (Fusco et al. 2004). The method has been proved to be quite robust in the past for differential imaging by NACO (Maire et al. 2014). We are therefore quite confident of our photometry corrected for the SR effect in J and Ks bands. 
To convert stellar fluxes into magnitudes, we set the zero-point magnitude to the instrumental zero-point (one ADU/s in J and Ks are 25.405 and 24.256 magnitudes, respectively) in addition to the SR effect {\footnote{
To verify our calibration method, we compared our WR stars K magnitude with six WR stars in Crowther et al. 2010. R136c is located farther from the very bright core of R136 and it has one close neighbor with a K-flux ratio of 0.114. R136c in Crowther et al. 2010 has Kmag of $11.34 \pm 0.08$, and in our catalog, it is 11.48, considering the flux of the companion in addition to R136c, then it becomes 11.36.}}.

In the second step, after extracting sources, the background image 
was used to estimate the residual errors in addition to the formal photometric PSF fitting errors of {\it{Starfinder}}. 
The background image contains 1) the AO halo from the atmospheric turbulence, 2) residuals from the photometric analysis in the first step and 3) undetected faint stars. 
We define the residual error as the fluctuation of the background caused by the remaining flux from the photometry using {\it{Starfinder}} and also from the undetected faint sources.

Since the core of R\,136 is crowded with the brightest stars, the AO halo is brightest at the center of R136. 
We removed this large halo in the Fourier space by applying a high pass filter (hat function with the diameter equal to the FWHM of the background halo), in order to estimate the fast variations of the background at the scale of the PSF. 
The residual error is the standard deviation of the background (after removing the stellar PSF and the AO halo) in the 30$\times$30 pixel$^2$ area around the source. 
The final photometric error is set to the quadratic combination of the PSF fitting errors, residual errors, and the photometric calibration error (0.042 in K and 0.128 in J).

\begin{figure}
\centering
\includegraphics[trim=0 0 0 0,clip,width=\textwidth/2]{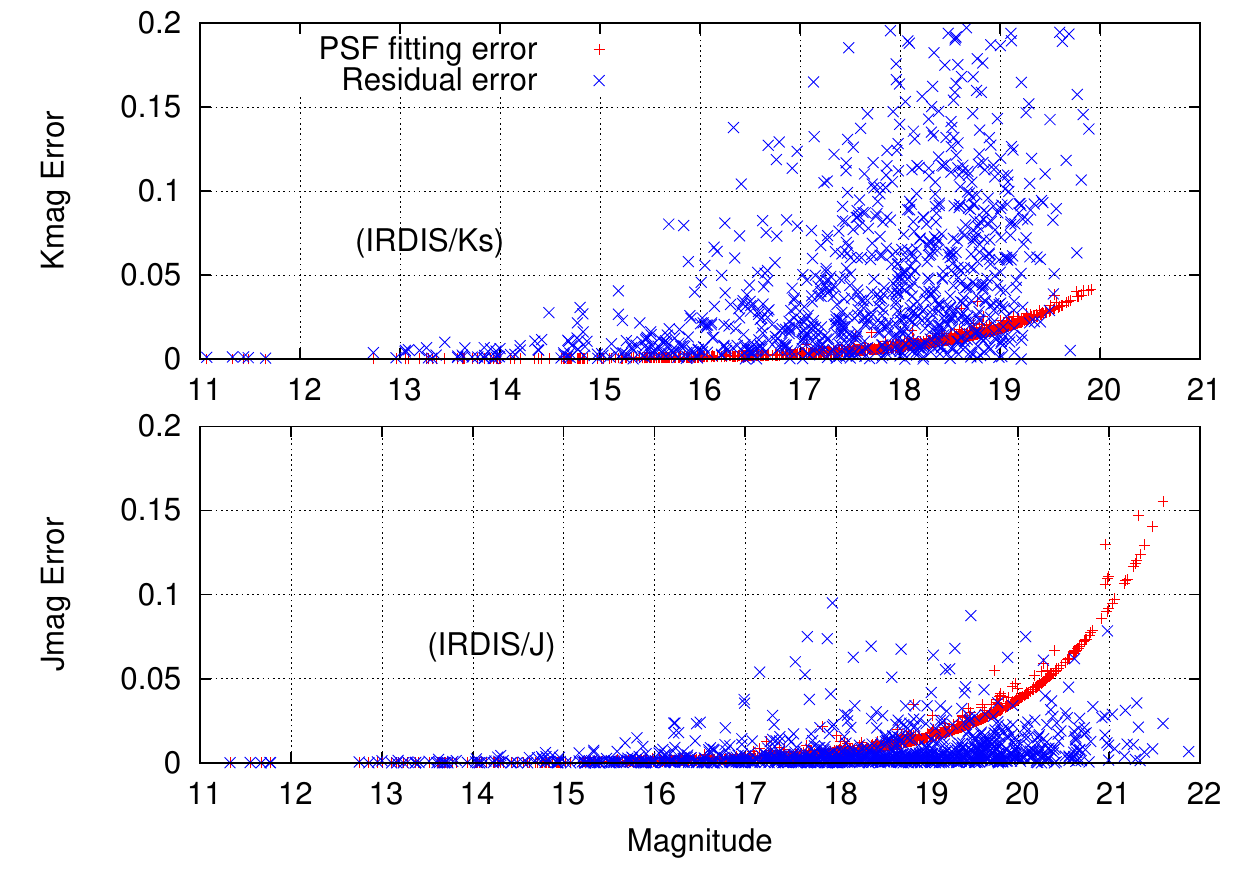}
\caption{PSF fitting errors (red pluses) and residual errors (blue crosses) in Ks (top) and J (bottom).
The PSF fitting error is the outcome of the {\it{Starfinder}}.
The residual errors are the outcome of the background analysis after removing the stellar sources from the image.}
\label{fig:2error}
\end{figure}

\begin{figure*}
	\centering
	\includegraphics[trim=30 0 30 0,clip,width=8.2cm]{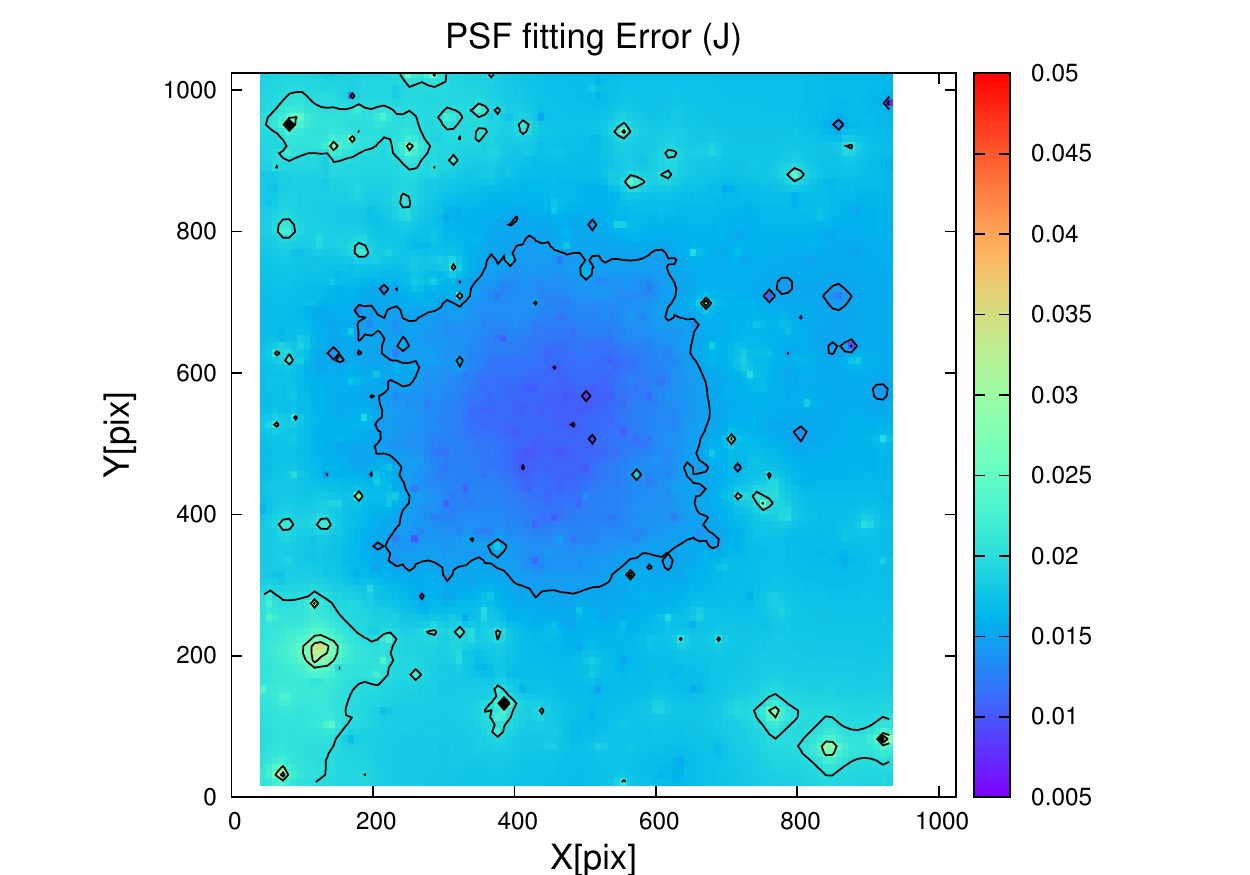}
	\includegraphics[trim=30 0 30 0,clip,width=8.2cm]{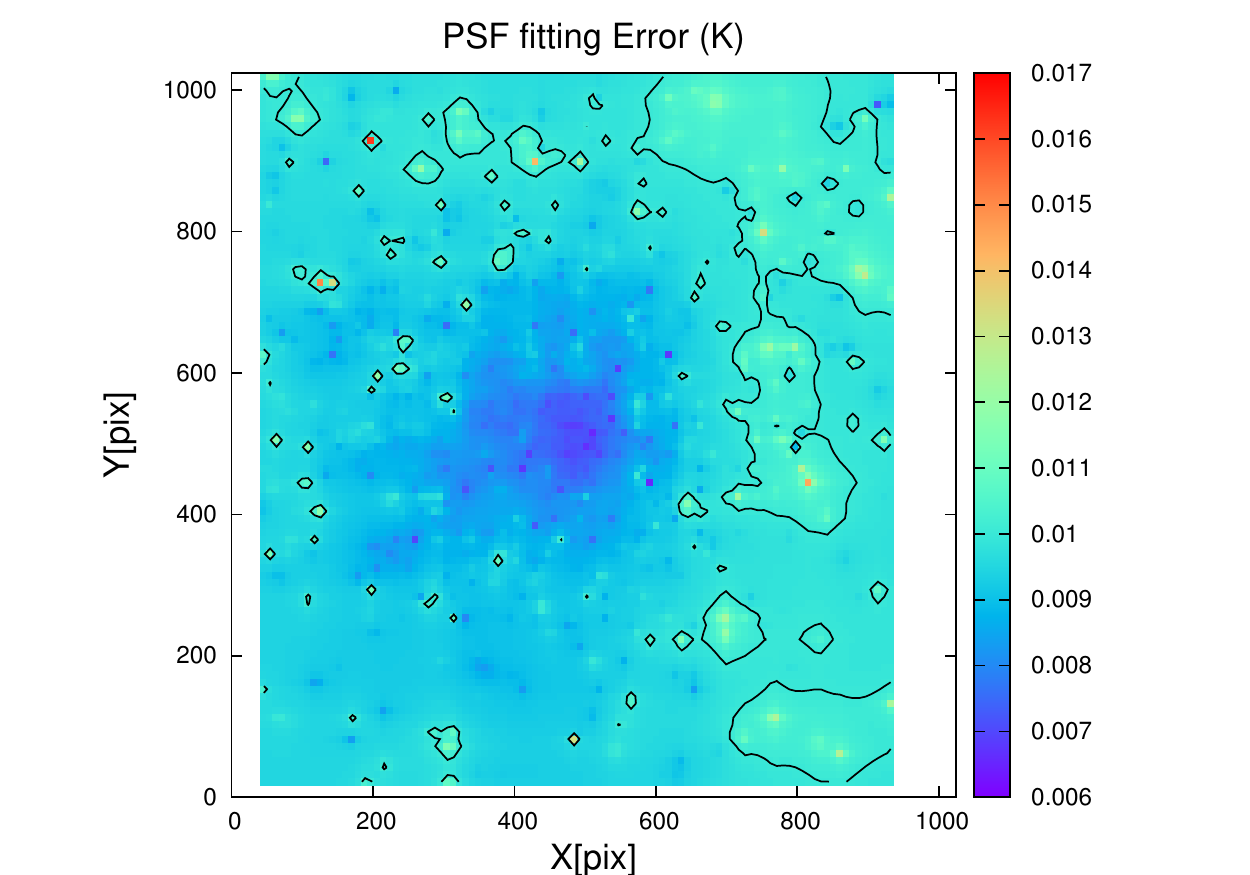}\\
	\includegraphics[trim=30 0 30 0,clip,width=8.2cm]{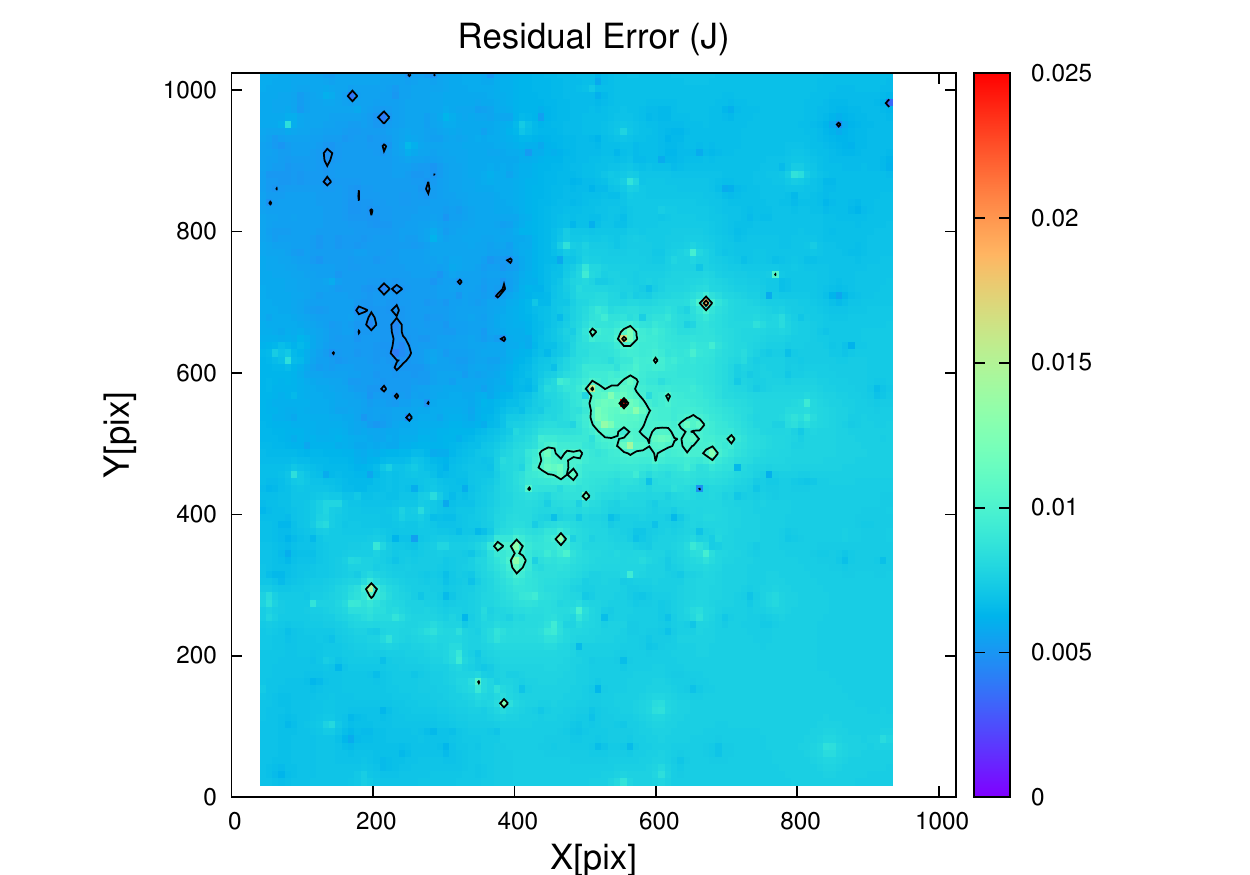}
	\includegraphics[trim=30 0 30 0,clip,width=8.2cm]{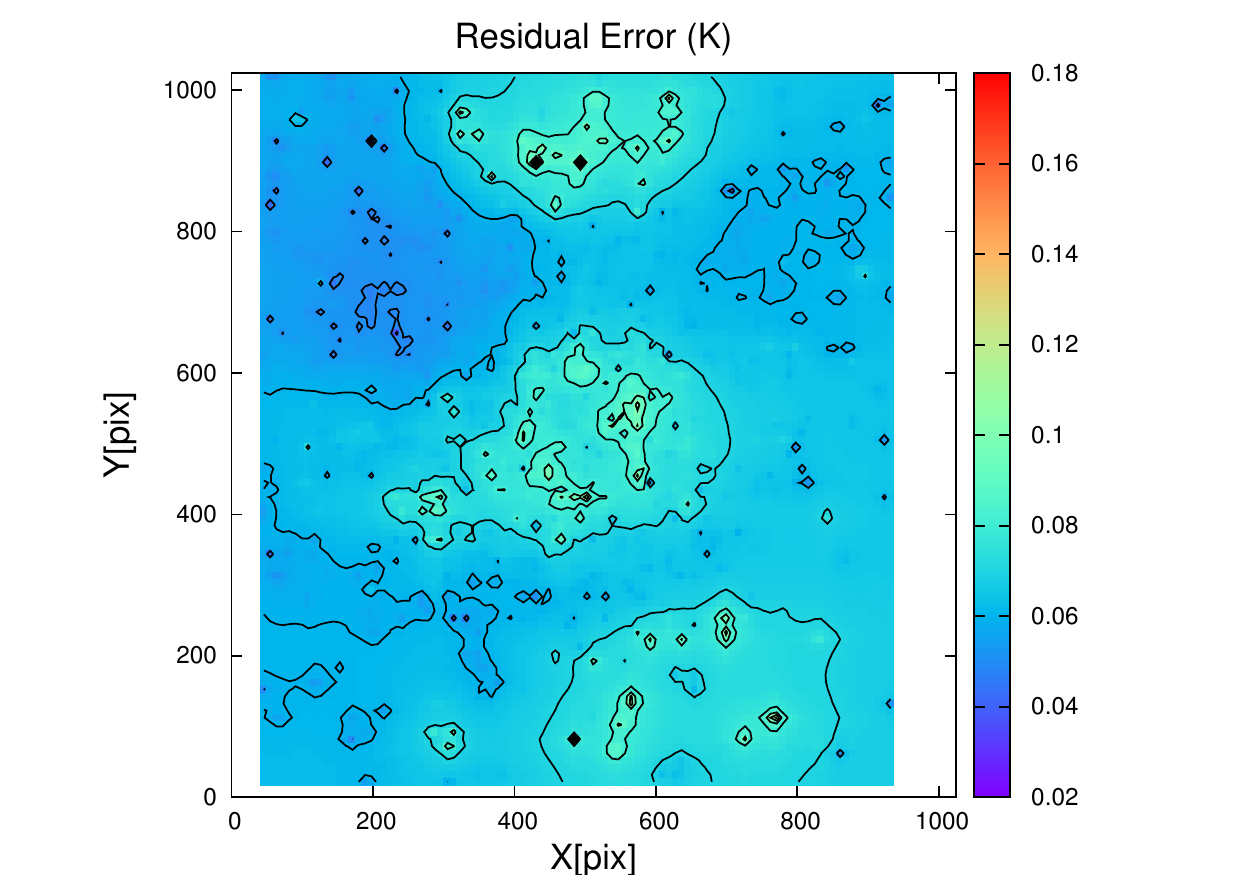}\\
	\includegraphics[trim=30 0 30 0,clip,width=8.2cm]{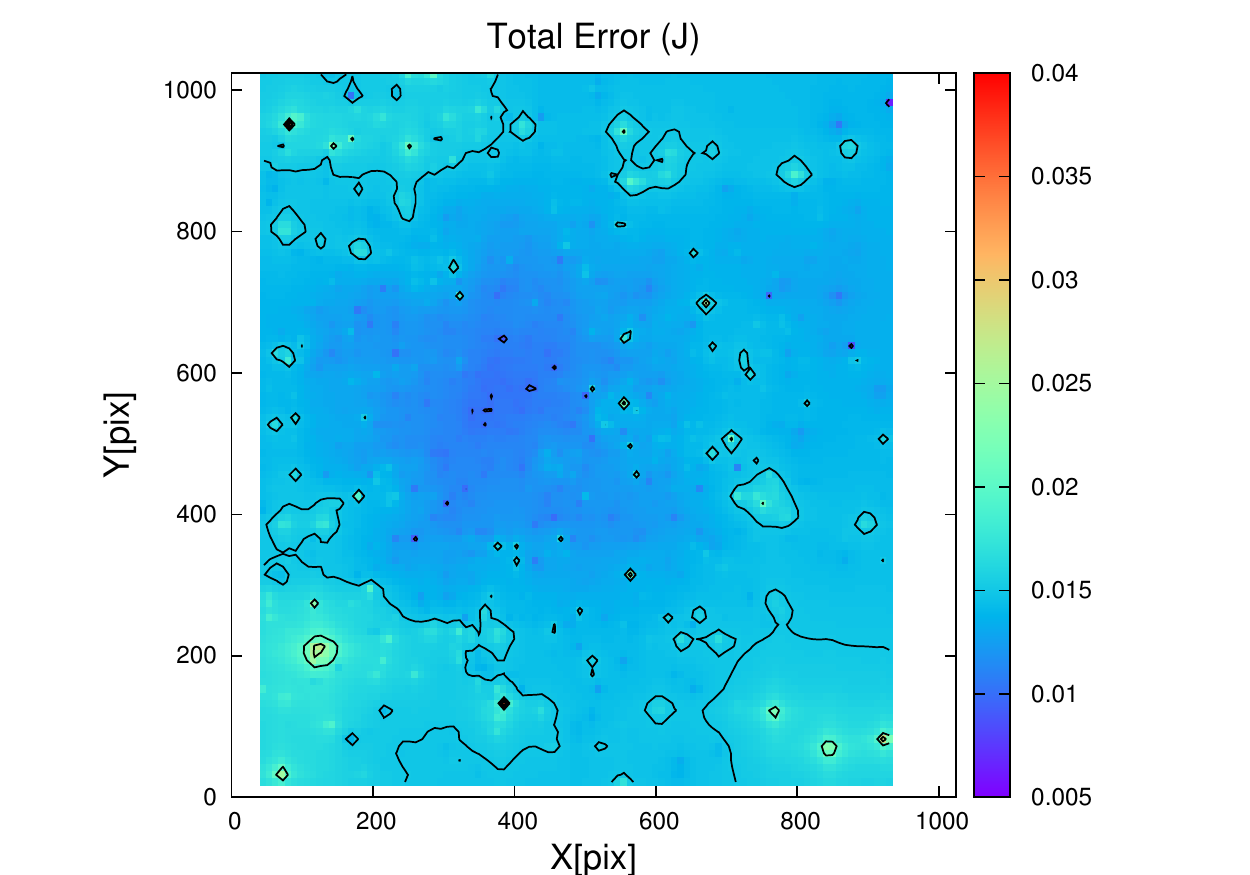}
	\includegraphics[trim=30 0 30 0,clip,width=8.2cm]{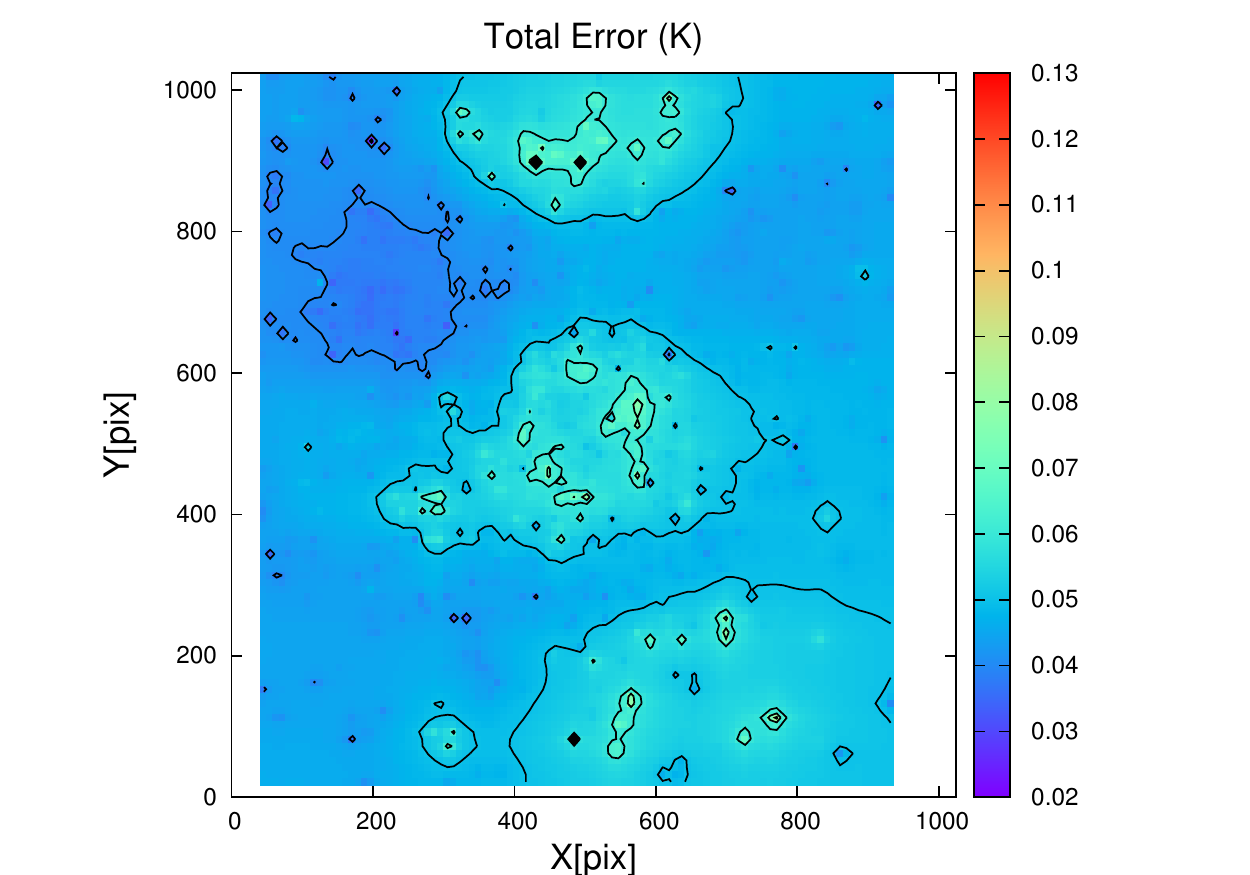}
	\caption{Map of the photometric, residual, and total relative error bars. Top: map of the PSF fitting errors (outcome pf the Satrfinder photometry) along the IRDIS FoV. 
		Middle: map of the residual errors, outcome of the background analysis after removing the stellar source signals from the image.
		Bottom: map of the total relative error, combination of the PSF fitting errors and the residual background errors.
		Left images are in the J band. Right images are in the Ks band.}
	\label{fig:map2error}
\end{figure*}

Figure \ref{fig:2error} shows these errors separately for each detected source.
The PSF fitting errors (red-pluses) are smaller in Ks band because the AO works better in longer wavelengths. The residual errors (blue crosses) are larger in Ks band, however, because the background fluctuations at longer wavelengths are higher.
Figure \ref{fig:map2error} shows the maps of these two errors in J and Ks across the field.

\begin{figure}
\centering
\includegraphics[trim=0 0 0 0,clip,width=\textwidth/2]{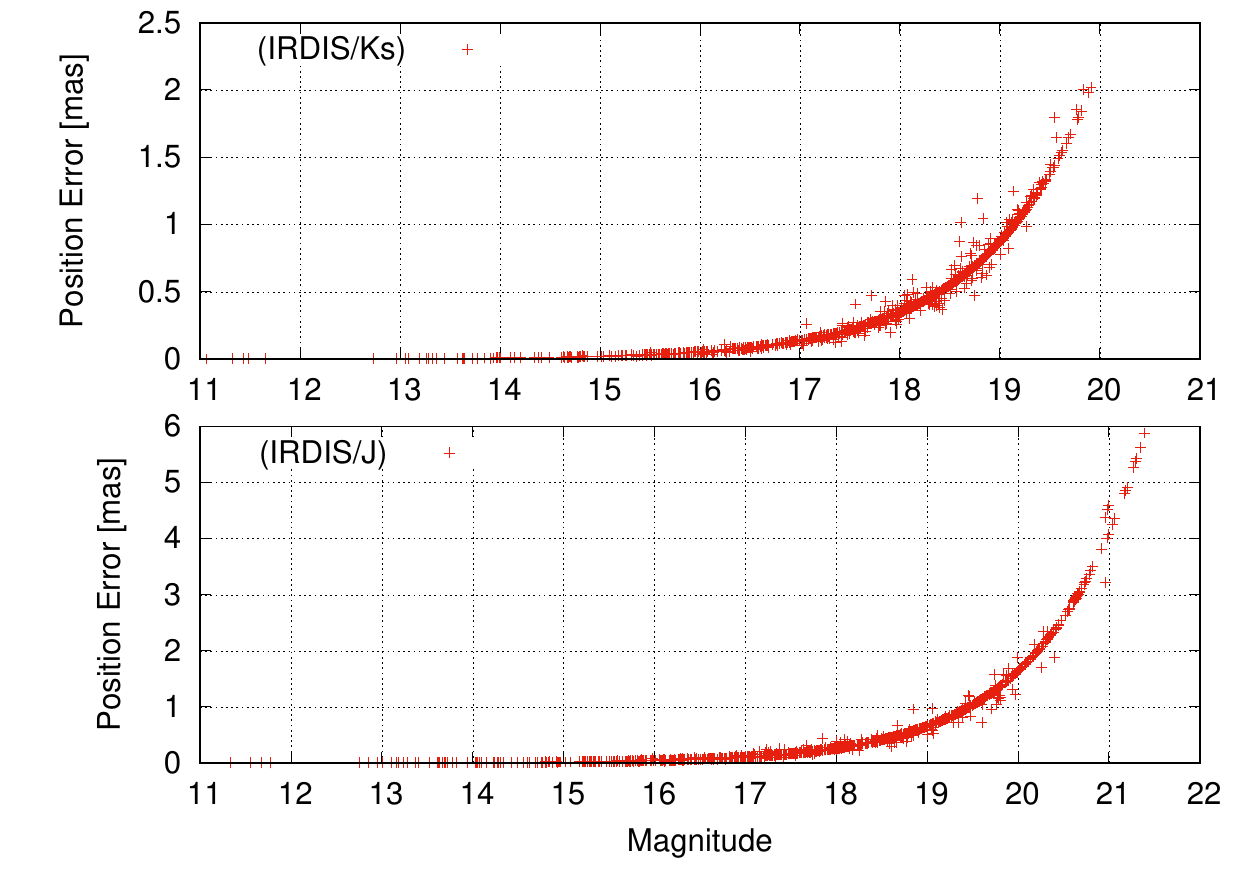}
\caption{Error in the position of the stars in Ks (top) and J (bottom) data.
This error is the outcome of the {\it{Starfinder}} and decreases for the bright stars.}
\label{fig:xy2error}
\end{figure}

The error on the position of the detected stars is estimated by {\it{Starfinder}} and shown in Figure \ref{fig:xy2error}. All stars brighter than 16.7 magnitude in Ks and 17 magnitude in J have position errors smaller than 0.1 mas (about 5 AU).

\begin{figure}
\centering
\includegraphics[trim=0 0 0 0,clip,width=\textwidth/2]{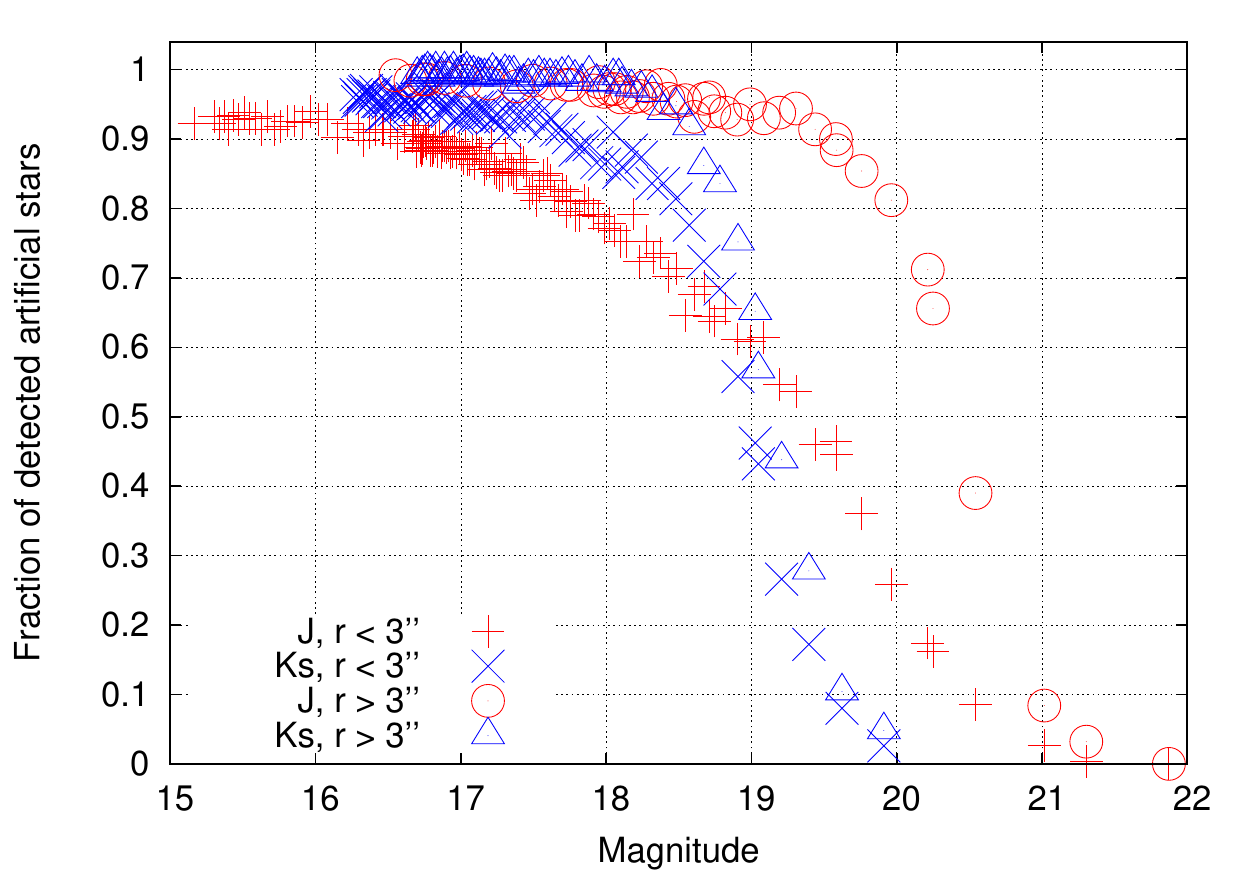}
\caption{Incompleteness test in J (red) and Ks (blue) for two regions: the very core of R\,136 (r < 3"), shown with pluses and crosses, and outside of the core (r > 3"), shown with circles and triangles. For each flux value (magnitude) we used 500 artificial stars in order to find the completeness value. Note that artificial stars are added one by one to the original image, and there is always one artificial star in each experiment.
}
\label{fig:incomjk}
\end{figure}

In order to interpret the photometric distribution of the 1110 and 1059 sources in J and Ks bands, we conducted an incompleteness test to the core (r < 3") of R\,136 and outside its core (r > 3") in both J- and Ks-band imaging data. 
Figure \ref{fig:incomjk} shows the result of incompleteness tests, which were performed by adding artificial stars with known magnitudes, one by one, to the original images. These experiments were repeated 500 times for each flux value (magnitude).

The faintest star in the artificial star test has a flux of 83.14 [ADU/s] in J band (Jmag=21.87) and 328.614 [ADU/s] in K -band (Kmag=19.91). For this star we iterated the test 500 times in the inner region (r < 3") and 500 times in the outer region ($r > 3"$). This experiment was made for 187 (in J) and 158 (in Ks) fluxes, 500 times for each flux value, with the $\Delta(flux)$ about 12.5 [ADU/s] on average.
Each time, one artificial star was added to these images and {\it{Starfinder}} was used again to detect the artificial star.
The core of the cluster is very crowded so that the incompleteness does not reach 100\% even for the bright artificial stars. 
This effect is more important for J-band data where the core is fuzzy because of a lower AO correction.

\begin{figure}
\centering
\includegraphics[trim=30 0 45 0,clip,width=8.2cm]{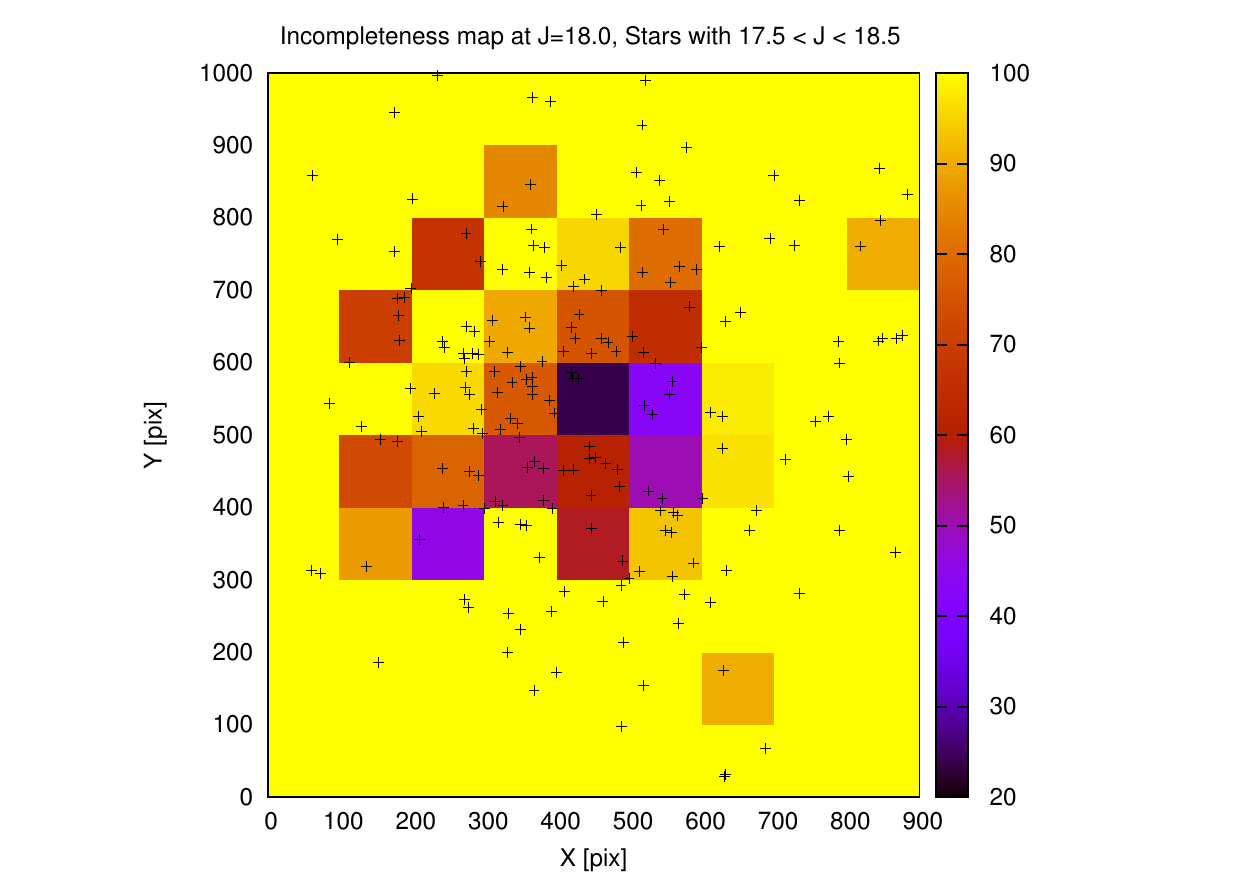}
\includegraphics[trim=30 0 45 0,clip,width=8.2cm]{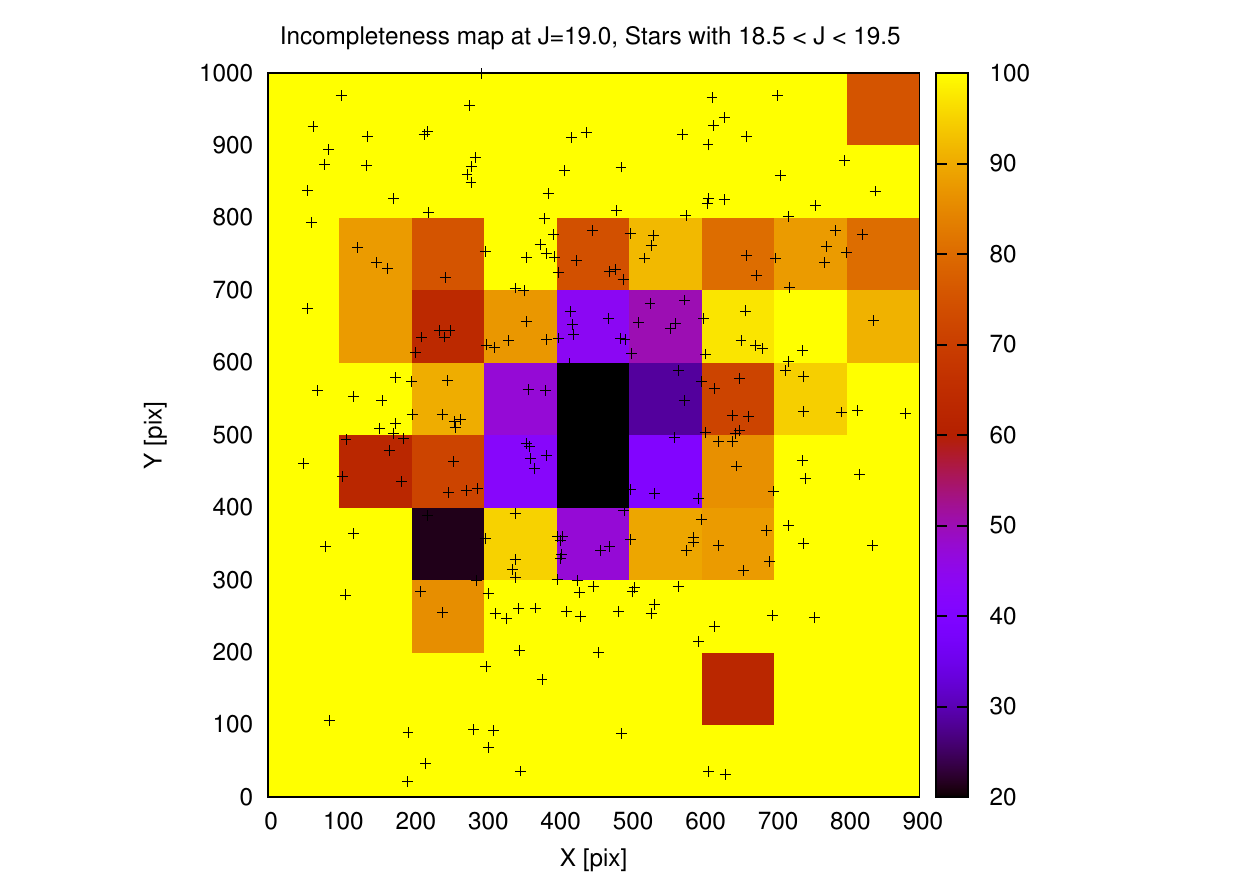}
\caption{Surface incompleteness map for magnitude 18 (top) and 19 (bottom) in J band. The incompleteness increases toward the central crowded core of R\,136 and also close to the bright stars. The black pluses indicate the position of the stars that are almost as bright as the artificial star test.}
\label{fig:surincom}
\end{figure}

\begin{figure*}
\centering
\includegraphics[trim=0 0 55 0,clip,width=14cm]{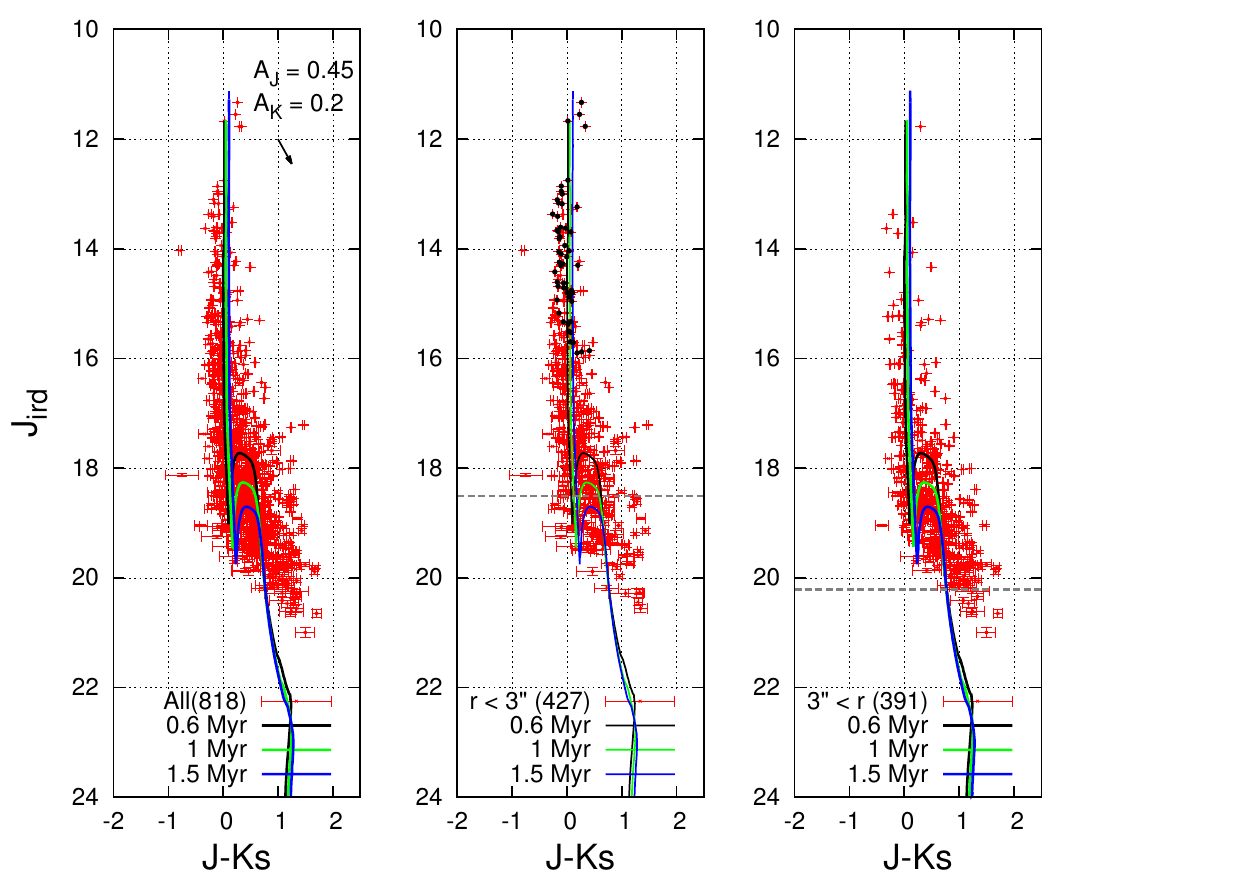}\\
\includegraphics[trim=0 0 5 0,clip,width=14cm]{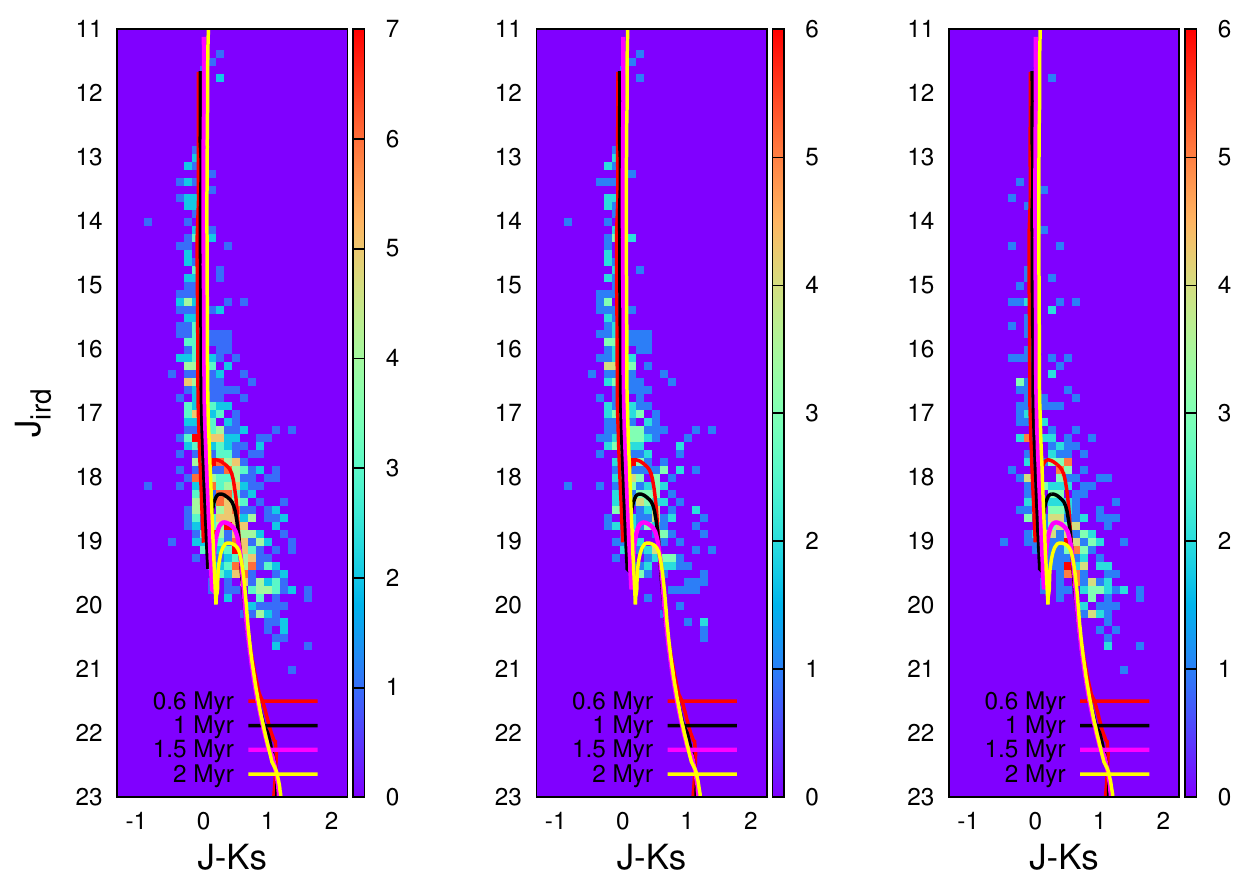}
\caption{Top: CMD of 818 detected sources in J- and Ks-band images of SPHERE/IRDIS from the core of R136. Solid black, green and blue lines show the PARSEC isochrones at the ages of 0.6, 1 and 1.5 Myr (corrected for distance modulus of 18.49 and central values of extinctions, $A_J =0.45~mag$ and $A_K=0.2~mag$). The horizontal dashed gray line depicts the incompleteness of 70\%.
The CMD is plotted for the whole FoV (818 sources), in the very core of the cluster (r < 3") and outside (r > 3"), from left to right respectively. The error bar on each point is the combination of the PSF fitting errors and the residual errors from the background image after removing the stellar source signals from the images. The 54 spectroscopically studied stars by Crowther et al. 2016 are shown as black dots in the middle plot.
Bottom: Same as top CMD, but in 3D to show the number density of stars in the CMD. Solid red, black, pink, and yellow lines depict PARSEC isochrones at the age of 0.5, 1, 1.5, and 2 Myr, respectively.}
\label{fig:cmderror}
\end{figure*}

\section{Age and extinction}
\begin{figure*}
\centering
\includegraphics[trim=0 0 0 1,clip,width=7cm]{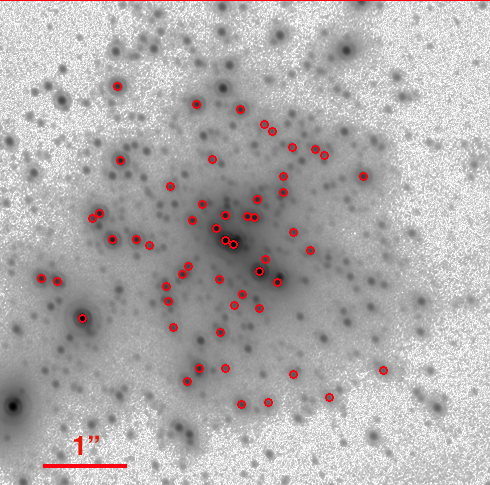}
\includegraphics[trim=0 10 12 5,clip,width=10.5cm]{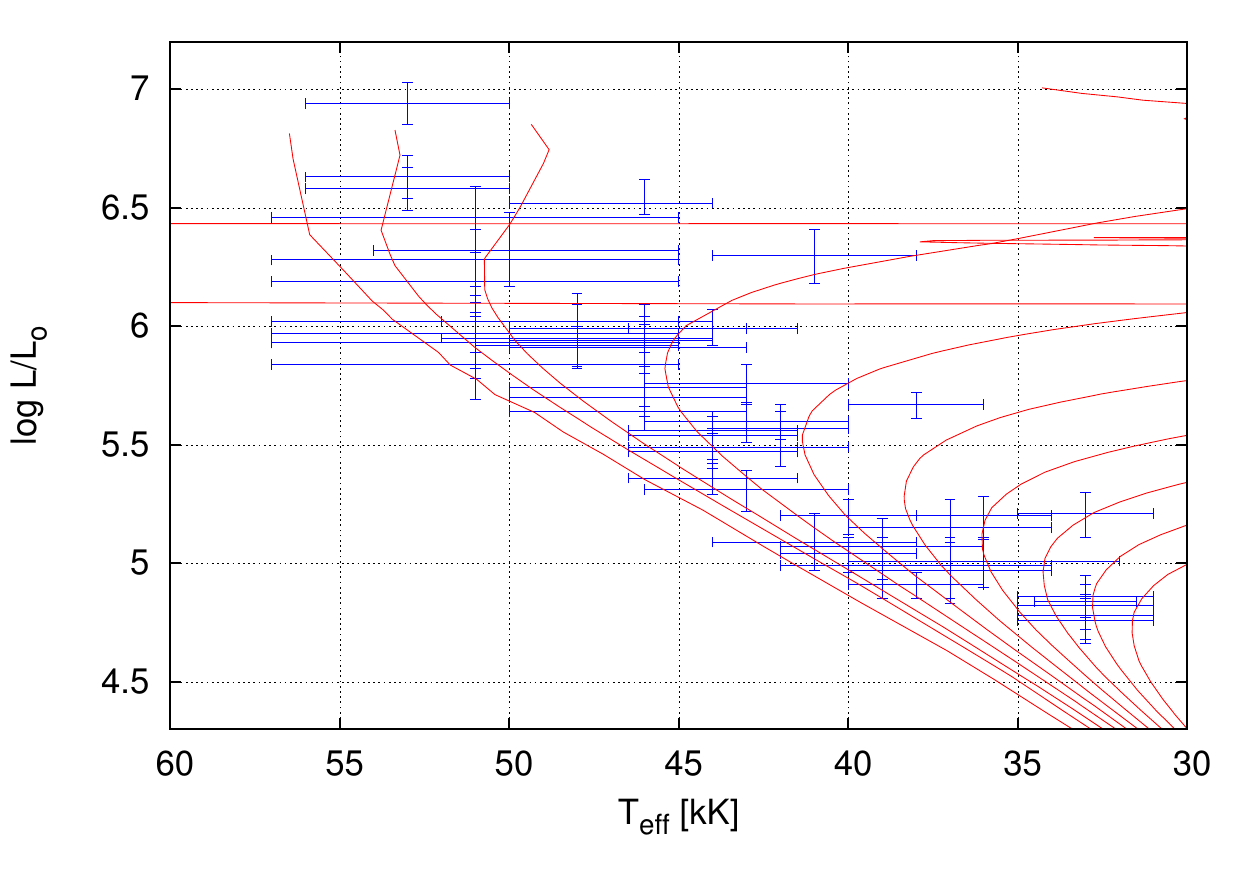}
\caption{Left panel: the IRDIS/Ks (FoV 6"$\times$6") on which the 54 spectroscopically known stars from Crowther et al. (2016) have been added as red circles.
The right plot depicts the $T_{eff}$, logL/L$_\odot$, and corresponding error bars on these 54 sources taken from Crowther et al. 2016 in blue.
The solid red lines indicate the PARSEC isochrones covering ages from 0.1 to 8 Myr.}
\label{fig:crowther55}
\end{figure*}

To estimate the stellar ages and the extinction of the core of R136, we used the effective temperature ($T_{eff}$) and luminosity (logL) of 54 stars that were studied spectroscopically by Crowther et al. (2016). 
We also chose a grid of isochrones at different ages (from 0.1 up to 8 Myr) with the LMC metallicity (Z=0.006) from the latest sets of PARSEC evolutionary model \footnote{http://stev.oapd.inaf.it/cgi-bin/cmd} (Bressan et al. 2012), which is a complete theoretical library that includes the latest set of stellar phases from pre-main sequence to main sequence and covering stellar masses from 0.1 to 350 M$_{\odot}$.
Figure \ref{fig:crowther55} shows these selected 54 stars with their $T_{eff}$-logL (with their error bars) and the sets of isochrones covering them.

By fitting the isochrones to each star, we estimated the age and intrinsic color of each star with the error bars. We adopted the LMC distance modulus (DM) of 18.49 magnitude estimated by Pietrzy{\'n}ski et al. (2013), which is consistent with the value suggested by Gibson (2000) for the LMC.

\begin{figure}
\centering
\includegraphics[trim=0 0 0 0,clip,width=8.cm]{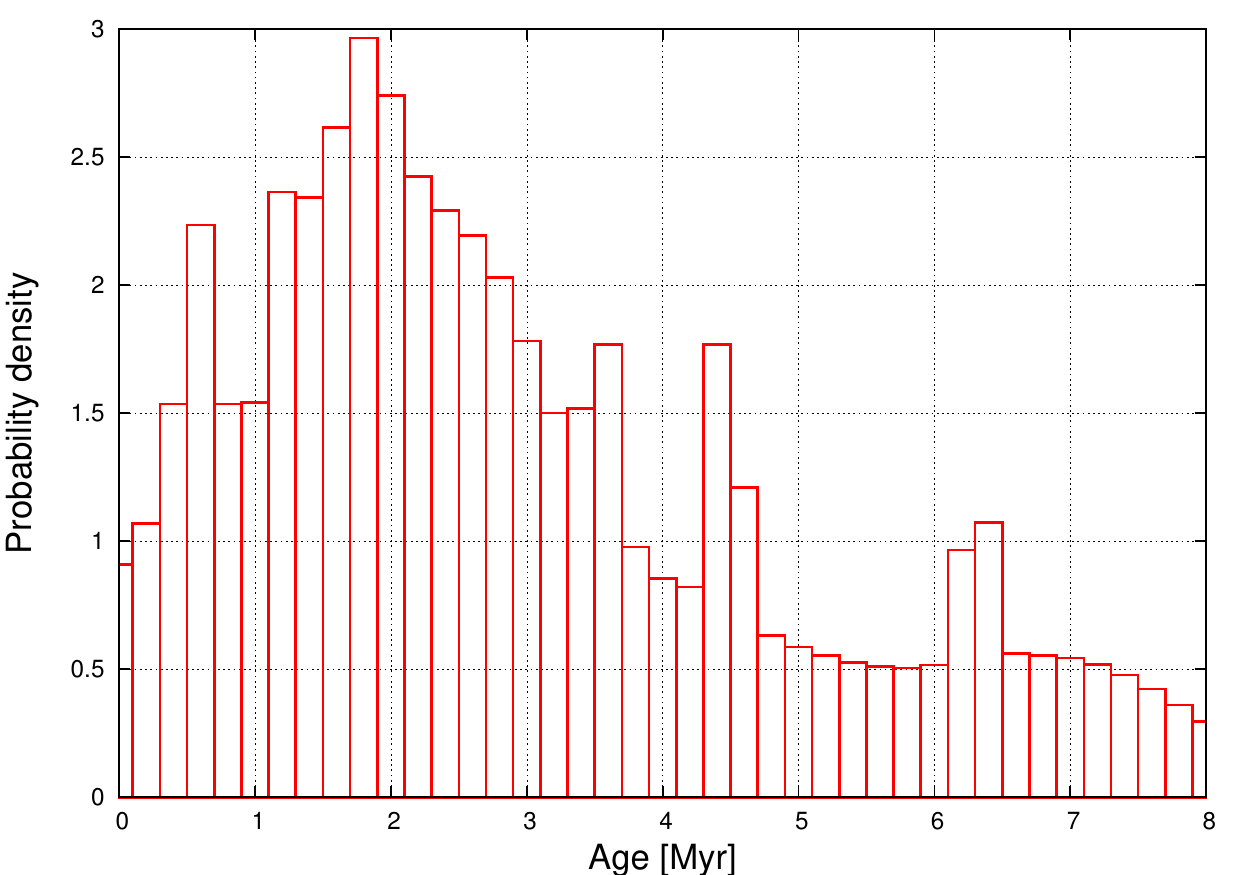}
\caption{Generalized histogram of the age of 54 known stars from Crowther et al. 2016 (shown in Figure \ref{fig:crowther55}). We used PARSEC isochrones at different ages to estimate the age-range for each spectroscopically known star.
This result can be compared with the age distribution of this sample that was observed in the UV by Crowther et al. 2016 (their Fig. 11).
}
\label{fig:histage}
\end{figure}

Table 2
shows the estimated age, initial and actual mass and extinction of these 54 spectroscopically known stars. The values of $T_{eff}$ and logL/L$_{\odot}$ in the second and third column are taken from Crowther et al. 2016.
Figure \ref{fig:histage} shows the generalized histogram of the age of these 54 sources. We note that the age of each star has a Gaussian distribution with a given $\sigma$ (error) in the histogram, derived from the errors in $T_{eff}$ and $L_*$.
We also note that the large errors on the age and extinction come from the large spectroscopic uncertainties (errors on $T_{eff}$ and logL/L$_\odot$). We were also limited by the evolutionary tracks up to 350 M$_{\odot}$, which explains the upper mass limit of 348 M$_\odot$ for very massive stars like R136a1.

\begin{figure}
\centering
\includegraphics[trim=0 0 0 0,clip,width=8.cm]{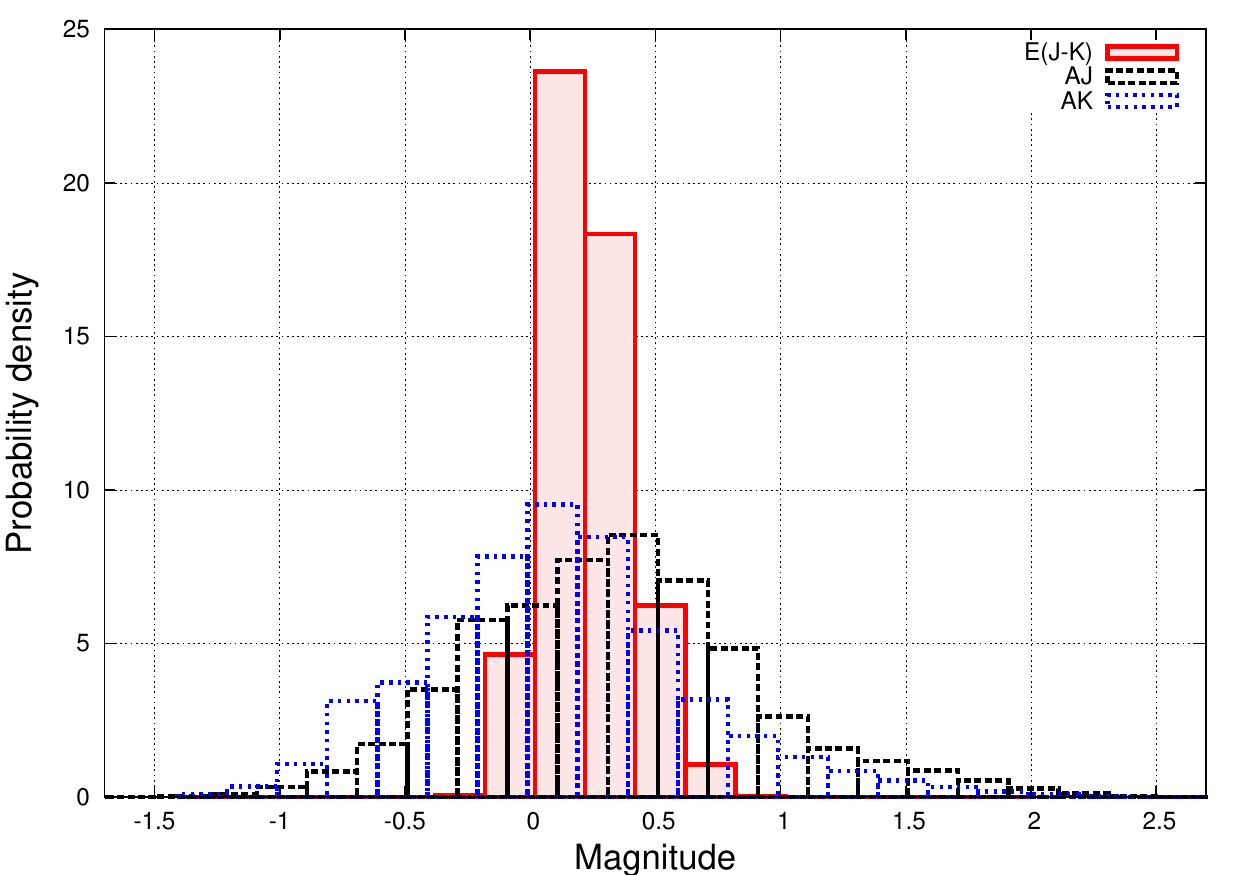}
\caption{Generalized histogram of the extinction of 54 spectroscopically known stars from Crowther et al. 2016 (shown in Figure \ref{fig:crowther55}). We used PARSEC models to estimate the extinction for each of these stars according to its age range.
 }
\label{fig:histextinction}
\end{figure}

Figure \ref{fig:histextinction} shows the histogram of the extinction in J and K and their color-excess.

The age of $1.8^{+1.2}_{-0.8}$ Myr is the most probable age range for these stars. 
Comparing Figure \ref{fig:histage} with Figure 11 of Crowther et al. (2016), the age estimate for this sample of stars is consistent within the error bars in both analyses.

The extinction in J and Ks is $0.45\pm0.5$ and $0.2\pm0.5$ magnitude, respectively. The estimated extinctions are consistent with the values derived in previous studies by Campbell et al. (2010) and De Marchi \& Panagia 2014. The color excess of $E(J-K)=0.25 \pm 0.1$ is also consistent with Tatton et al. (2013).
Figure \ref{fig:cmderror} shows the color-magnitude diagram (CMD) of detected sources in J- and Ks-band IRDIS data with their error bars. The CMD is plotted for the whole FoV (818 sources), in the very core of the cluster (r < 3"), and outside (r > 3"), from left to right. The error bars on each point are the combination of the PSF fitting errors and the residual errors from the background image after removing the stellar source signals from the images.
The PARSEC isochrones at three different ages (0.6, 1, and 1.5 Myr) also are plotted in this figure using DM=18.49 and central extinction values in J (0.45 mag) and K (0.2 mag).

We note that based on the CMD isochrone fitting, the age of 1 and 1.5 Myr for the inner ($r < 3"$) and outside core ($r > 3 "$) fits best to our data. We can also see a young population with an age of about 0.6 Myr in both regions. In the histogram of the estimated age for the 54 spectroscopically known stars there is a peak at 0.6 Myr that shows a probable age for some stars. 
We can clearly see the main sequence and pre-main sequence conjunction, which is very sensitive to age. The age and extinction can be estimated using the precise photometric analysis for the 818 stellar sources in the core of R136.
Interestingly, the age of stars in the very core ($r < 3"$) is younger than in its outer region ($r > 3"$). 
This result appears to be consistent with the older population found beyond the core of R\,136, like the northeast clump observed by Sabbi et al. (2012) and at a greater distance (3'), the old cluster, Hodge\,301, studied by Grebel \& Chu (2000). Understanding this apparent age trend in R\,136 and also 30\,Dor region is possible in the future from the formation point of view. Moreover, the question is how the younger population in the center of the cluster can be explained by star cluster formation scenarios (W{\"u}nsch et al. 2017)?
We note that this age difference in the two regions can also be explained by an observational bias because the central region of R\,136 is very compact and bright so that the incompleteness level is very low. Figure \ref{fig:surincom} shows the incompleteness map at J=18 (top) and J=19 (bottom). The incompleteness is low in the very central regions, which means that we did not detect the true number of stars. This might affect the CMD in this magnitude range, which is sensitive to age.

\begin{figure}
\centering
\includegraphics[trim=3 5 5 5,clip,width=8.2cm]{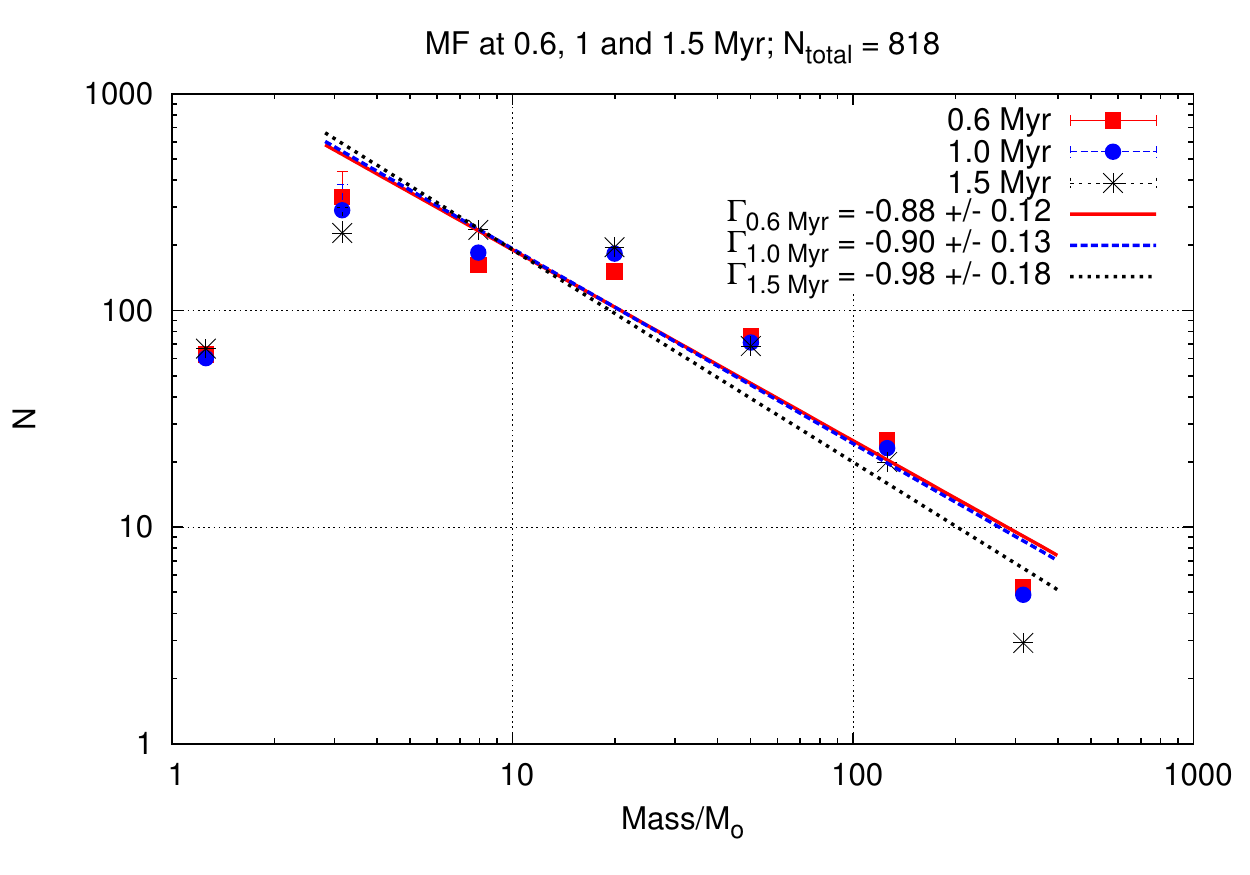}
\caption{Generalized histogram of the stellar masses (MF) at 0.6, 1 and 1.5 Myr. PARSEC models used to estimate the stellar-mass range for each source using extinction-range.}
\label{fig:histmf}
\end{figure}

Considering the errors on the extinction, we can estimate the stellar mass range for each star at a given age. 
We estimated the stellar masses only for common sources between J and Ks data (818 sources shown in Fig. \ref{fig:cmderror}) using their J and Ks magnitudes fitted to PARSEC isochrones at three different ages: 0.6, 1, and 1.5 My.
The histogram of mass, which is the mass function (MF), is plotted considering a Gaussian distribution for each stellar mass. 
The Gaussian uncertainty in the mass of each star was accounted for when constructing the MF.
Figure \ref{fig:histmf} shows the generalized histogram of the  mass (MF) at three different ages (0.6, 1, and 1.5 Myr). 
The MF slope for 1 and 1.5 Myr isochrone is $\Gamma_{1Myr}=-0.90\pm0.13$ and $\Gamma_{1.5Myr}=-0.98\pm0.18$, respectively, for the mass range of (3 - 300) M$_{\odot}$. These values are low limits to steepness because of incompleteness and central concentration of bright stars.
These slopes{\footnote{See Table 5 and Sect. 7 for the comparison with previous works.}} are flatter than the Kroupa ($\Gamma=-1.3$, Kroupa (2001)) and Salpeter value ($\Gamma=-1.35$, Salpeter(1955)).
We note that the incompleteness correction is applied for the low-mass bin considering the number of stars in the inner ($r < 3"$) and outer ($r > 3"$) region of the cluster (see Figure \ref{fig:incomjk}). 
The derived MF is limited to the resolution of the instrument and also to the detection limit of the observation. In future, using higher angular resolution data, we may resolve binaries and low-mass stars, which affect the shape of MF.

\begin{table*} 	\label{table:54starinfo}  
\begin{center}
\caption{Information about 54 spectroscopically known stars with $T_{eff}$ and logL/L$_{\odot}$ estimated by Crowther et al. 2016 (second and third columns). 
Using PARSEC evolutionary isochrones (0.1 to 8 Myr), we estimated the age, color excess, and initial and actual stellar masses (columns four, five, eight and nine). IRDIS Ks and J magnitudes are given in column six and seven.
 N2 and N1 are the number of visual companions for each source in a radius of 0.2" and 0.1", respectively.
The identifications (ID) of the sources are from Hunter et al. 1995. Note that we are also limited by the evolutionary tracks up to 350 M$_{\odot}$ ,which explains zero error bars for very bright stars.}
\begin{tabular}{c c c c c c c c c c c}
HSH95&$T_{eff}$[kK]&logL/L$_{\odot}$&age[Myr]&E(J-K)&Ks&J&$M_{initial}$&$M_{act}$&N2&N1\\
\hline
  3&$53.00^{+3.0}_{-3.0}$  &$6.94^{+0.09}_{-0.09 }$  &$0.79^{+1.44}_{-0.00}$  &$ 0.49^{+0.02}_{-0.00}$  &$ 11.07\pm0.04$&$ 11.33\pm0.13$&$348.1^{+ 0.0}_{-80.1}$&$325.2^{+ 0.0}_{-147.0}$&3&1\\
5&$53.00^{+3.0}_{-3.0}$  &$6.63^{+0.09}_{-0.09 }$  &$0.56^{+0.00}_{-0.36}$  &$ 0.46^{+0.00}_{-0.00}$  &$ 11.32\pm0.04$&$ 11.55\pm0.13$&$201.5^{+48.5}_{- 0.0}$&$195.5^{+51.7}_{-  0.0}$&3&0\\
20&$50.00^{+4.0}_{-5.0}$  &$6.32^{+0.16}_{-0.15 }$  &$1.26^{+2.29}_{-0.86}$  &$ 0.27^{+0.01}_{-0.02}$  &$ 12.73\pm0.04$&$ 12.75\pm0.13$&$120.0^{+30.2}_{-47.9}$&$114.1^{+32.5}_{- 72.6}$&2&0\\
24&$46.00^{+4.0}_{-3.0}$  &$5.99^{+0.10}_{-0.08 }$  &$1.78^{+2.69}_{-0.78}$  &$ 0.14^{+0.01}_{-0.01}$  &$ 13.05\pm0.04$&$ 12.95\pm0.13$&$ 75.4^{+14.6}_{-22.2}$&$ 72.2^{+15.1}_{- 44.4}$&3&0\\
27&$51.00^{+6.0}_{-6.0}$  &$6.28^{+0.13}_{-0.15 }$  &$1.00^{+2.55}_{-0.90}$  &$ 0.15^{+0.02}_{-0.02}$  &$ 13.27\pm0.04$&$ 13.18\pm0.13$&$120.0^{+30.0}_{-47.9}$&$115.6^{+33.8}_{- 74.1}$&1&1\\
21&$51.00^{+6.0}_{-6.0}$  &$6.46^{+0.13}_{-0.15 }$  &$1.12^{+1.70}_{-1.02}$  &$ 0.17^{+0.01}_{-0.03}$  &$ 13.09\pm0.04$&$ 13.00\pm0.13$&$150.2^{+39.9}_{-29.3}$&$142.1^{+35.2}_{- 61.0}$&1&1\\
86&$46.00^{+4.0}_{-3.0}$  &$5.91^{+0.10}_{-0.08 }$  &$1.78^{+2.69}_{-0.98}$  &$ 0.25^{+0.04}_{-0.04}$  &$ 14.80\pm0.05$&$ 14.80\pm0.13$&$ 67.9^{+11.1}_{-14.8}$&$ 65.4^{+10.9}_{- 37.7}$&2&0\\
66&$46.00^{+4.0}_{-3.0}$  &$5.70^{+0.10}_{-0.08 }$  &$1.78^{+0.73}_{-1.68}$  &$ 0.14^{+0.02}_{-0.02}$  &$ 14.42\pm0.04$&$ 14.31\pm0.13$&$ 53.7^{+10.2}_{- 4.8}$&$ 52.4^{+10.7}_{-  4.8}$&2&0\\
6&$53.00^{+3.0}_{-3.0}$  &$6.58^{+0.09}_{-0.09 }$  &$0.56^{+0.23}_{-0.00}$  &$ 0.57^{+0.00}_{-0.00}$  &$ 11.44\pm0.04$&$ 11.77\pm0.13$&$201.5^{+ 0.0}_{-24.4}$&$195.5^{+ 0.0}_{- 25.5}$&2&1\\
58&$51.00^{+6.0}_{-6.0}$  &$5.93^{+0.13}_{-0.15 }$  &$0.79^{+3.67}_{-0.69}$  &$ 0.12^{+0.01}_{-0.02}$  &$ 13.91\pm0.04$&$ 13.78\pm0.13$&$ 75.0^{+15.1}_{-21.7}$&$ 73.8^{+16.0}_{- 46.5}$&0&0\\
70&$42.00^{+2.0}_{-2.0}$  &$5.57^{+0.07}_{-0.08 }$  &$2.51^{+0.65}_{-0.27}$  &$ 0.10^{+0.01}_{-0.01}$  &$ 14.20\pm0.04$&$ 14.06\pm0.13$&$ 44.2^{+ 1.7}_{- 4.2}$&$ 43.1^{+ 1.5}_{-  4.2}$&1&1\\
30&$38.00^{+2.0}_{-2.0}$  &$5.67^{+0.05}_{-0.06 }$  &$3.55^{+0.00}_{-0.39}$  &$ 0.21^{+0.01}_{-0.01}$  &$ 13.65\pm0.04$&$ 13.62\pm0.13$&$ 45.0^{+ 3.1}_{- 0.4}$&$ 43.0^{+ 3.0}_{-  0.3}$&2&0\\
89&$44.00^{+2.5}_{-2.5}$  &$5.99^{+0.08}_{-0.07 }$  &$4.47^{+0.00}_{-2.69}$  &$ 0.18^{+0.01}_{-0.01}$  &$ 14.69\pm0.04$&$ 14.62\pm0.13$&$ 53.1^{+23.0}_{- 0.0}$&$ 27.8^{+44.4}_{-  0.0}$&2&0\\
62&$51.00^{+6.0}_{-6.0}$  &$5.84^{+0.13}_{-0.15 }$  &$0.20^{+1.80}_{-0.10}$  &$ 0.16^{+0.03}_{-0.02}$  &$ 14.35\pm0.04$&$ 14.28\pm0.13$&$ 70.0^{+ 5.0}_{-16.3}$&$ 69.8^{+ 5.1}_{- 17.4}$&4&0\\
19&$46.00^{+4.0}_{-2.0}$  &$6.52^{+0.10}_{-0.05 }$  &$1.26^{+0.00}_{-0.26}$  &$ 0.43^{+0.01}_{-0.00}$  &$ 13.05\pm0.04$&$ 13.24\pm0.13$&$170.0^{+20.2}_{- 1.8}$&$157.9^{+19.5}_{-  0.0}$&2&1\\
50&$51.00^{+6.0}_{-6.0}$  &$6.02^{+0.15}_{-0.13 }$  &$4.47^{+0.00}_{-4.37}$  &$ 0.10^{+0.02}_{-0.02}$  &$ 13.96\pm0.04$&$ 13.82\pm0.13$&$ 53.3^{+46.7}_{- 0.0}$&$ 27.4^{+72.4}_{-  0.0}$&2&0\\
90&$44.00^{+2.5}_{-2.5}$  &$5.36^{+0.08}_{-0.07 }$  &$0.79^{+1.72}_{-0.69}$  &$ 0.29^{+0.02}_{-0.02}$  &$ 14.83\pm0.05$&$ 14.88\pm0.13$&$ 38.6^{+ 1.4}_{- 4.6}$&$ 38.4^{+ 1.6}_{-  4.9}$&1&0\\
141&$41.00^{+3.0}_{-3.0}$  &$5.09^{+0.12}_{-0.12 }$  &$2.00^{+1.99}_{-1.90}$  &$ 0.18^{+0.03}_{-0.03}$  &$ 15.40\pm0.05$&$ 15.33\pm0.13$&$ 28.3^{+ 5.6}_{- 3.4}$&$ 28.1^{+ 5.7}_{-  3.4}$&4&0\\
80&$36.00^{+2.0}_{-2.0}$  &$5.20^{+0.08}_{-0.10 }$  &$4.47^{+1.16}_{-0.00}$  &$ 0.18^{+0.01}_{-0.01}$  &$ 14.77\pm0.04$&$ 14.71\pm0.13$&$ 28.0^{+ 0.7}_{- 3.5}$&$ 27.6^{+ 0.6}_{-  3.4}$&1&0\\
35&$48.00^{+3.0}_{-3.0}$  &$5.92^{+0.08}_{-0.09 }$  &$1.58^{+0.41}_{-1.19}$  &$ 0.29^{+0.01}_{-0.02}$  &$ 14.00\pm0.04$&$ 14.04\pm0.13$&$ 70.0^{+ 5.4}_{- 8.3}$&$ 67.7^{+ 6.5}_{-  8.3}$&2&0\\
78&$44.00^{+2.5}_{-2.5}$  &$5.47^{+0.08}_{-0.07 }$  &$2.24^{+0.58}_{-1.68}$  &$ 0.32^{+0.01}_{-0.01}$  &$ 14.68\pm0.04$&$ 14.76\pm0.13$&$ 40.0^{+ 5.0}_{- 2.2}$&$ 39.3^{+ 5.5}_{-  2.3}$&0&0\\
73&$33.00^{+2.0}_{-2.0}$  &$5.21^{+0.09}_{-0.10 }$  &$6.31^{+0.00}_{-0.69}$  &$ 0.09^{+0.01}_{-0.01}$  &$ 14.38\pm0.04$&$ 14.24\pm0.13$&$ 25.4^{+ 2.6}_{- 1.4}$&$ 24.9^{+ 2.5}_{-  1.3}$&2&0\\
92&$40.00^{+2.0}_{-2.0}$  &$5.20^{+0.07}_{-0.08 }$  &$2.82^{+1.16}_{-1.56}$  &$ 0.33^{+0.01}_{-0.02}$  &$ 14.87\pm0.04$&$ 14.96\pm0.13$&$ 30.0^{+ 2.1}_{- 2.4}$&$ 29.7^{+ 2.2}_{-  2.4}$&1&0\\
143&$39.00^{+3.0}_{-3.0}$  &$4.99^{+0.12}_{-0.14 }$  &$3.16^{+1.85}_{-3.06}$  &$ 0.06^{+0.02}_{-0.01}$  &$ 15.11\pm0.04$&$ 14.93\pm0.13$&$ 24.9^{+ 5.1}_{- 3.3}$&$ 24.7^{+ 5.2}_{-  3.3}$&4&0\\
112&$36.00^{+4.0}_{-4.0}$  &$5.01^{+0.10}_{-0.11 }$  &$4.47^{+2.61}_{-3.47}$  &$ 0.06^{+0.01}_{-0.01}$  &$ 14.78\pm0.04$&$ 14.61\pm0.13$&$ 24.0^{+ 4.0}_{- 4.0}$&$ 23.8^{+ 4.0}_{-  4.0}$&3&1\\
135&$33.00^{+2.0}_{-2.0}$  &$4.86^{+0.09}_{-0.10 }$  &$7.08^{+0.86}_{-2.07}$  &$ 0.09^{+0.01}_{-0.01}$  &$ 15.32\pm0.04$&$ 15.17\pm0.13$&$ 19.4^{+ 1.9}_{- 1.4}$&$ 19.3^{+ 1.8}_{-  1.4}$&3&1\\
69&$42.00^{+2.0}_{-2.0}$  &$5.49^{+0.07}_{-0.08 }$  &$2.82^{+0.34}_{-0.82}$  &$ 0.14^{+0.01}_{-0.01}$  &$ 14.39\pm0.04$&$ 14.29\pm0.13$&$ 40.0^{+ 1.4}_{- 2.2}$&$ 39.1^{+ 1.6}_{-  2.1}$&3&0\\
52&$46.00^{+4.0}_{-3.0}$  &$5.74^{+0.10}_{-0.08 }$  &$2.00^{+0.52}_{-1.80}$  &$ 0.14^{+0.01}_{-0.01}$  &$ 14.20\pm0.04$&$ 14.10\pm0.13$&$ 55.0^{+10.0}_{- 5.0}$&$ 53.4^{+11.1}_{-  5.0}$&2&0\\
48&$51.00^{+6.0}_{-6.0}$  &$5.97^{+0.13}_{-0.15 }$  &$1.26^{+3.21}_{-1.16}$  &$ 0.08^{+0.00}_{-0.02}$  &$ 13.58\pm0.04$&$ 13.41\pm0.13$&$ 75.0^{+20.0}_{-21.7}$&$ 73.0^{+21.8}_{- 45.6}$&3&0\\
94&$43.00^{+3.0}_{-3.0}$  &$5.31^{+0.08}_{-0.09 }$  &$2.00^{+1.17}_{-1.90}$  &$ 0.08^{+0.01}_{-0.01}$  &$ 14.85\pm0.04$&$ 14.69\pm0.13$&$ 34.8^{+ 5.2}_{- 3.5}$&$ 34.5^{+ 5.5}_{-  3.4}$&2&0\\
115&$33.00^{+2.0}_{-2.0}$  &$4.82^{+0.09}_{-0.10 }$  &$7.08^{+0.86}_{-2.07}$  &$ 0.30^{+0.01}_{-0.02}$  &$ 15.46\pm0.04$&$ 15.53\pm0.13$&$ 18.8^{+ 1.5}_{- 1.0}$&$ 18.7^{+ 1.5}_{-  1.0}$&1&0\\
132&$33.00^{+2.0}_{-2.0}$  &$4.76^{+0.09}_{-0.10 }$  &$7.08^{+0.86}_{-2.61}$  &$ 0.25^{+0.01}_{-0.01}$  &$ 15.36\pm0.04$&$ 15.38\pm0.13$&$ 18.1^{+ 2.2}_{- 0.9}$&$ 18.0^{+ 2.1}_{-  0.8}$&2&0\\
36&$46.00^{+4.0}_{-2.0}$  &$5.94^{+0.10}_{-0.05 }$  &$1.78^{+0.22}_{-0.78}$  &$ 0.14^{+0.00}_{-0.00}$  &$ 12.96\pm0.04$&$ 12.86\pm0.13$&$ 70.0^{+12.9}_{- 2.1}$&$ 67.3^{+13.0}_{-  2.1}$&1&0\\
173&$33.00^{+2.0}_{-2.0}$  &$4.78^{+0.09}_{-0.10 }$  &$6.31^{+1.63}_{-1.84}$  &$ 0.30^{+0.01}_{-0.02}$  &$ 15.63\pm0.04$&$ 15.70\pm0.13$&$ 19.2^{+ 1.1}_{- 1.9}$&$ 19.1^{+ 1.1}_{-  1.9}$&3&0\\
75&$44.00^{+2.5}_{-2.5}$  &$5.54^{+0.08}_{-0.07 }$  &$1.58^{+1.23}_{-1.02}$  &$ 0.31^{+0.01}_{-0.01}$  &$ 14.78\pm0.04$&$ 14.84\pm0.13$&$ 45.0^{+ 1.7}_{- 5.0}$&$ 44.4^{+ 1.5}_{-  5.3}$&2&0\\
114&$37.00^{+3.0}_{-3.0}$  &$4.99^{+0.12}_{-0.14 }$  &$3.98^{+1.64}_{-3.19}$  &$ 0.30^{+0.01}_{-0.01}$  &$ 15.26\pm0.04$&$ 15.33\pm0.13$&$ 24.0^{+ 4.0}_{- 2.7}$&$ 23.8^{+ 4.0}_{-  2.7}$&1&0\\
108&$37.00^{+3.0}_{-3.0}$  &$5.15^{+0.12}_{-0.14 }$  &$3.98^{+1.64}_{-1.47}$  &$ 0.35^{+0.02}_{-0.01}$  &$ 15.59\pm0.04$&$ 15.71\pm0.13$&$ 27.6^{+ 2.6}_{- 4.5}$&$ 27.3^{+ 2.6}_{-  4.4}$&2&0\\
31&$51.00^{+6.0}_{-6.0}$  &$6.19^{+0.13}_{-0.15 }$  &$1.26^{+2.72}_{-1.16}$  &$ 0.32^{+0.00}_{-0.02}$  &$ 13.62\pm0.04$&$ 13.69\pm0.13$&$100.0^{+21.1}_{-38.8}$&$ 96.0^{+23.6}_{- 61.7}$&1&0\\
49&$43.00^{+3.0}_{-3.0}$  &$5.60^{+0.08}_{-0.09 }$  &$2.51^{+0.65}_{-1.39}$  &$ 0.20^{+0.01}_{-0.00}$  &$ 13.98\pm0.04$&$ 13.94\pm0.13$&$ 45.0^{+ 5.0}_{- 5.0}$&$ 43.8^{+ 5.1}_{-  4.9}$&1&0\\
46&$48.00^{+4.0}_{-4.0}$  &$6.02^{+0.12}_{-0.09 }$  &$1.26^{+3.21}_{-0.70}$  &$ 0.07^{+0.00}_{-0.02}$  &$ 13.29\pm0.04$&$ 13.11\pm0.13$&$ 82.9^{+12.1}_{-29.6}$&$ 80.3^{+12.6}_{- 53.0}$&3&0\\
47&$48.00^{+4.0}_{-4.0}$  &$5.95^{+0.14}_{-0.13 }$  &$1.41^{+3.05}_{-1.31}$  &$ 0.09^{+0.00}_{-0.02}$  &$ 13.32\pm0.04$&$ 13.17\pm0.13$&$ 73.5^{+16.6}_{-20.2}$&$ 71.3^{+17.6}_{- 43.9}$&3&0\\
40&$51.00^{+6.0}_{-6.0}$  &$5.97^{+0.13}_{-0.15 }$  &$1.26^{+3.21}_{-1.16}$  &$ 0.13^{+0.00}_{-0.02}$  &$ 13.72\pm0.04$&$ 13.61\pm0.13$&$ 75.0^{+20.0}_{-21.7}$&$ 73.0^{+21.8}_{- 45.6}$&1&0\\
116&$37.00^{+3.0}_{-3.0}$  &$4.97^{+0.12}_{-0.14 }$  &$3.55^{+2.08}_{-3.45}$  &$ 0.65^{+0.01}_{-0.01}$  &$ 15.45\pm0.04$&$ 15.86\pm0.13$&$ 24.0^{+ 3.1}_{- 3.8}$&$ 23.9^{+ 3.1}_{-  3.7}$&1&1\\
118&$39.00^{+3.0}_{-3.0}$  &$5.07^{+0.12}_{-0.14 }$  &$3.16^{+1.85}_{-3.06}$  &$ 0.02^{+0.01}_{-0.00}$  &$ 14.64\pm0.04$&$ 14.42\pm0.13$&$ 26.5^{+ 4.0}_{- 3.3}$&$ 26.3^{+ 4.1}_{-  3.2}$&2&0\\
42&$46.00^{+4.0}_{-3.0}$  &$5.64^{+0.10}_{-0.08 }$  &$1.58^{+0.93}_{-1.48}$  &$ 0.44^{+0.01}_{-0.01}$  &$ 14.11\pm0.04$&$ 14.30\pm0.13$&$ 50.0^{+10.0}_{- 5.8}$&$ 49.1^{+10.7}_{-  6.0}$&0&0\\
55&$51.00^{+6.0}_{-6.0}$  &$5.93^{+0.13}_{-0.15 }$  &$0.79^{+3.67}_{-0.69}$  &$ 0.24^{+0.01}_{-0.02}$  &$ 14.15\pm0.04$&$ 14.15\pm0.13$&$ 75.0^{+15.1}_{-21.7}$&$ 73.8^{+16.0}_{- 46.5}$&0&0\\
71&$44.00^{+2.5}_{-2.5}$  &$5.49^{+0.08}_{-0.07 }$  &$1.78^{+1.04}_{-1.22}$  &$ 0.35^{+0.01}_{-0.01}$  &$ 14.70\pm0.04$&$ 14.80\pm0.13$&$ 42.3^{+ 2.7}_{- 4.5}$&$ 41.7^{+ 3.1}_{-  4.6}$&1&0\\
121&$33.00^{+1.5}_{-1.5}$  &$4.84^{+0.07}_{-0.07 }$  &$7.08^{+0.86}_{-1.46}$  &$ 0.27^{+0.01}_{-0.01}$  &$ 15.47\pm0.04$&$ 15.50\pm0.13$&$ 19.4^{+ 0.6}_{- 1.4}$&$ 19.3^{+ 0.6}_{-  1.4}$&1&0\\
9&$41.00^{+3.0}_{-3.0}$  &$6.30^{+0.11}_{-0.12 }$  &$1.78^{+0.22}_{-0.19}$  &$ 0.26^{+0.00}_{-0.01}$  &$ 11.66\pm0.04$&$ 11.67\pm0.13$&$119.6^{+17.7}_{-24.6}$&$109.9^{+16.4}_{- 21.7}$&1&0\\
65&$44.00^{+2.5}_{-2.5}$  &$5.56^{+0.08}_{-0.07 }$  &$2.00^{+0.82}_{-1.20}$  &$ 0.06^{+0.00}_{-0.01}$  &$ 13.85\pm0.04$&$ 13.67\pm0.13$&$ 45.0^{+ 5.0}_{- 5.0}$&$ 44.2^{+ 5.1}_{-  5.1}$&1&0\\
134&$38.00^{+2.0}_{-2.0}$  &$4.91^{+0.05}_{-0.06 }$  &$2.24^{+2.23}_{-1.44}$  &$ 0.42^{+0.01}_{-0.01}$  &$ 15.71\pm0.04$&$ 15.90\pm0.13$&$ 24.0^{+ 1.9}_{- 2.5}$&$ 23.9^{+ 1.9}_{-  2.5}$&1&0\\
64&$42.00^{+2.0}_{-2.0}$  &$5.60^{+0.07}_{-0.08 }$  &$2.51^{+0.65}_{-0.27}$  &$ 0.09^{+0.00}_{-0.00}$  &$ 13.84\pm0.04$&$ 13.69\pm0.13$&$ 45.0^{+ 1.6}_{- 5.0}$&$ 43.8^{+ 1.2}_{-  4.9}$&0&0\\
45&$43.00^{+3.0}_{-3.0}$  &$5.76^{+0.08}_{-0.09 }$  &$2.24^{+0.58}_{-0.46}$  &$-0.02^{+0.00}_{-0.00}$  &$ 13.63\pm0.04$&$ 13.37\pm0.13$&$ 55.0^{+ 5.0}_{- 5.0}$&$ 53.1^{+ 4.8}_{-  5.1}$&1&0\\
123&$40.00^{+2.0}_{-2.0}$  &$5.04^{+0.07}_{-0.08 }$  &$1.12^{+2.43}_{-1.02}$  &$ 0.51^{+0.01}_{-0.01}$  &$ 15.61\pm0.04$&$ 15.88\pm0.13$&$ 28.0^{+ 2.0}_{- 3.2}$&$ 27.9^{+ 2.0}_{-  3.2}$&0&0
\end{tabular}
\end{center}
\end{table*}

\section{Visual companions}
\begin{figure}
\centering
\includegraphics[trim=0 0 0 0,clip,width=8.2cm]{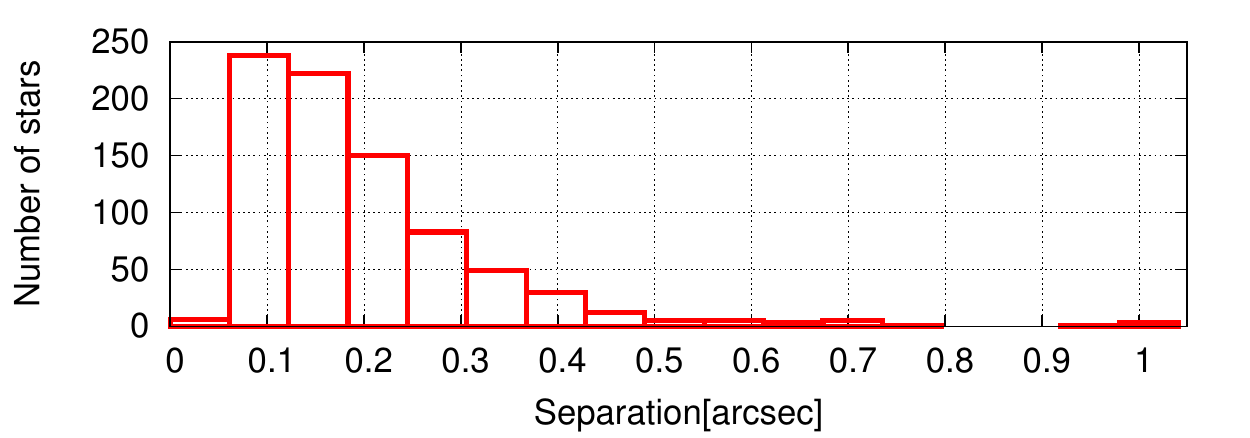}\\
\includegraphics[trim=0 0 0 0,clip,width=8.2cm]{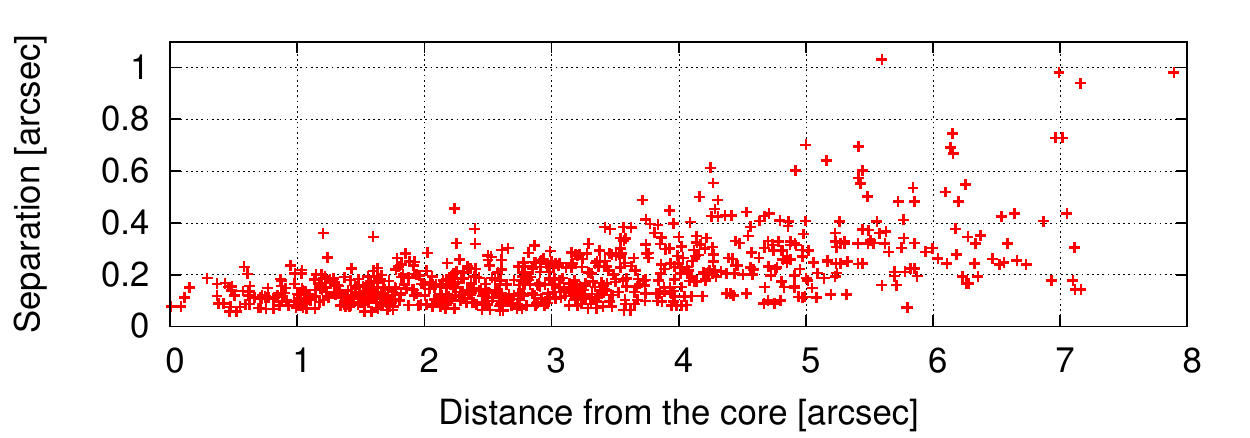}
\caption{Top: Histogram of the separation of the close detected sources. For each star that is detected in both J and K data, we determined a distance between the star and its closest neighbor.
Bottom: Separation of the visual close detected sources versus their distance from the core of R136.
}
\label{fig:close}
\end{figure}

For each star detected in both J and Ks, we determined the distance between the star and its closest neighbor. Figure \ref{fig:close} shows the number of closeby detected stars in Ks (red) and J (blue) versus their separation in arcseconds. 
More than 250 star pairs have a closest neighbor at a separation smaller than 0.2". 
Over 90\% of massive objects (brighter than 16.7 mag in Ks and 15 mag in J) have a closest neighbor with a separation smaller than 0.2". 
Figure \ref{fig:close} shows the separation between the visual close stars versus their distance from R136 in the core. This figure indicates that even the sources at larger radii have close visual companions, so that the large number of close visual companions in not just an effect of 2D projection on the sky across the FoV.
For the sake of simplicity, regardless of whether the stars are physically bound, we call these closeby stars {\bf{"visual companions"}} hereafter.

The most massive stars R136a1, R136a3, and R136c have visual companions. 
R136a3 is resolved as two stars with the PSF fitting. Both stars have a high correlation coefficient (above 70\%) with the input PSF. The separation between the R136a3 primary and secondary is about $58.9\pm 2.14 ~mas$, which is larger than the FWHM of the PSF. 
We note that even the closest visual companions (like R136a3) are physically far from each other (0.059" is 2890 AU). This visual separation produces a period over $P=10^4 ~yr$, so that these sources are probably not gravitationally bound to each other.
Table \ref{table:closestarinfo} shows the list of the 20 stars detected in both J and Ks data that have visual companions closer than 0.08". The flux ratio between two companions in Ks and J band are given in the third and fourth column, respectively. Their separation [in mas] is also given in the last column.

\begin{table}
\caption{List of 20 stars that have a companion closer than 0.08". These stars are detected in both J- and Ks-band data.} 
\centering
\begin{tabular}{c c c c}  
ID1,~ID2&$FluxRatio_{K}$&$FluxRatio_{J}$&Separation\\
&&&[mas]\\
\hline
~~~~~59,~3(a3)&$0.044\pm0.004$&$0.124\pm0.032$&$58.88\pm2.14$\\
~~357,~272&$0.699\pm0.032$&$0.577\pm0.023$&$59.87\pm0.17$\\
~~760,~519&$0.567\pm0.095$&$0.896\pm0.230$&$62.62\pm5.71$\\
~~643,~214&$0.208\pm0.020$&$0.139\pm0.036$ &$63.67\pm0.31$\\
319,~79& $0.166\pm0.018$&$0.177\pm0.045$&$64.11\pm4.23$\\
~~804,~517&$0.512\pm0.081$&$0.365\pm0.094$ &$65.06\pm1.40$\\
~~380,~265& $0.630\pm0.140$&$0.324\pm0.087$&$66.29\pm1.61$\\
~~807,~565&$0.571\pm0.072$&$0.830\pm0.228$ &$67.12\pm3.18$\\
11(a9),~8(a6)& $0.878\pm0.075$&$0.959\pm0.245$&$70.07\pm0.88$\\
~~~~25,~4(c)&$0.114\pm0.010$&$0.095\pm0.024$ &$70.27\pm0.07$\\
396,~56& $0.079\pm0.010$&$0.065\pm0.017$&$71.25\pm2.45$\\
~~635,~589&$0.870\pm0.126$&$0.730\pm0.192$ &$72.21\pm2.30$\\
~~541,~489&$0.867\pm0.079$&$0.964\pm0.247$ &$73.65\pm4.21$\\
~~637,~539&$0.774\pm0.067$&$1.002\pm0.258$ &$73.84\pm1.08$\\
312,~87&$0.197\pm0.019$&$0.188\pm0.048$ &$74.13\pm2.37$\\
~~72,~54&$0.686\pm0.066$&$0.714\pm0.183$ &$74.47\pm0.45$\\
~~~~~~16,~1(a1)& $0.112\pm0.010$&$0.152\pm0.039$&$75.32\pm2.33$\\
~~761,~655& $0.796\pm0.164$&$0.689\pm0.180$&$75.65\pm0.61$\\
~~353,~305& $0.819\pm0.096$&$0.450\pm0.115$&$75.91\pm0.94$\\
~~317,~262&$0.766\pm0.085$&$0.413\pm0.106$&$79.30\pm0.09$
\end{tabular} 
\label{table:closestarinfo} 
\end{table}

\section{Density and surface brightness profile}\label{sec:sbp}
\begin{figure}
\centering
\includegraphics[trim=0 0 0 0,clip,width=8.2cm]{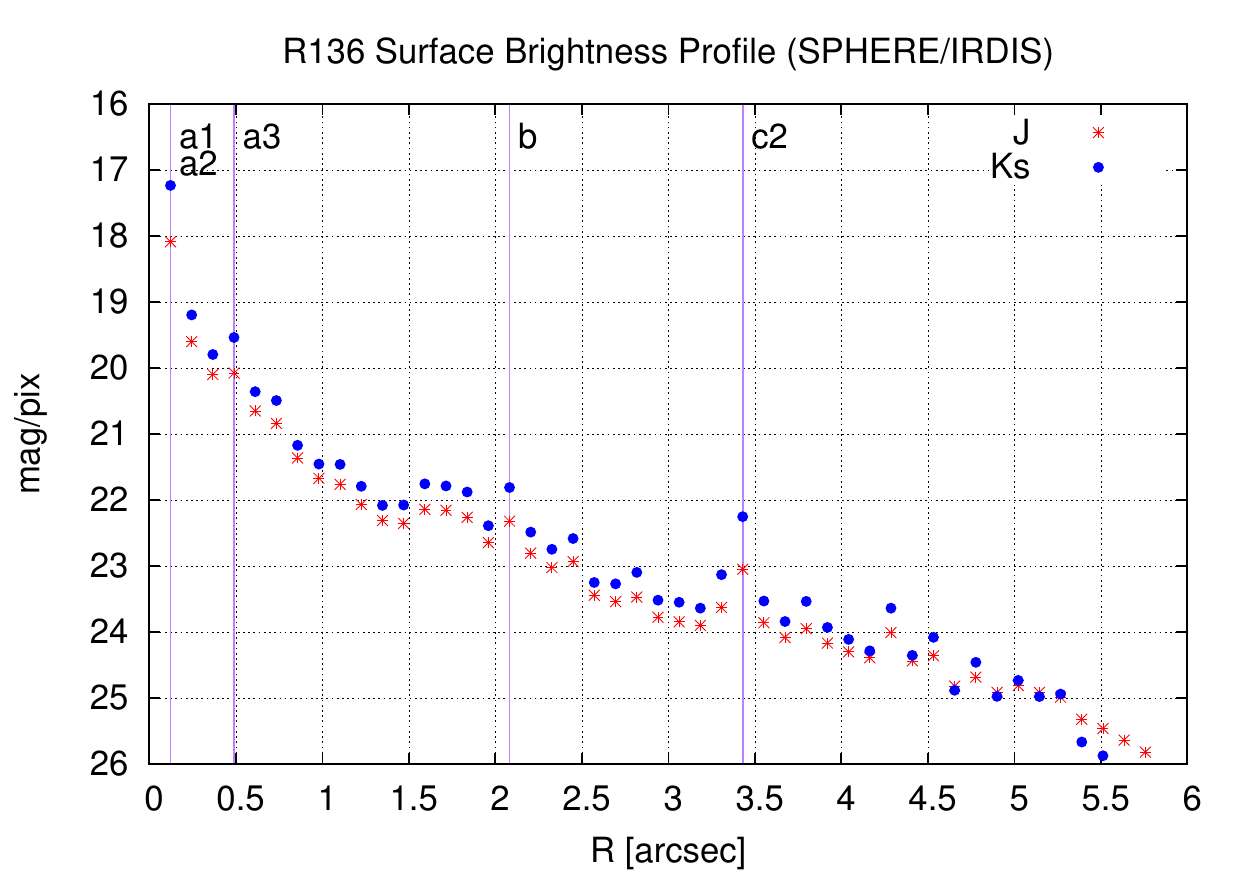}\\
\caption{SBP (mag/pixel) of R\,136 in IRDIS FoV in IRDIS FoV centered on R136a1. The radial positions of the five WR stars are shown with the solid vertical purple lines.
}
\label{fig:sbp}
\end{figure}

The unexpected number of detected sources in a small FoV and of resolved companions in R\,136 indicates that this compact cluster is more crowded than thought before.
The error bars on the stellar masses and also on the age and extinction of the cluster itself are large enough to make it difficult to study the density profile of R136. 
Instead, we can scrutinize the surface brightness profile (SBP) of this cluster, which is less affected by the confusion and crowding. We obtained the SBP by measuring radial profiles for J and Ks images centered on R136a1. The SBP informs us on the average magnitude per pixel at different radii. Figure \ref{fig:sbp} depicts the SBP of the core of R\,136 in J and K. 
On average, the SBP in Ks is brighter than J, which can be caused by the extinction or brighter stars in Ks. 
Several bumps on the SBP are visible at approximately 0.08, 0.15, 0.46, 1.17 pc radii.  
These radial distances are the locations of known WR stars. The position of five WRs in the FoV  is shown in the SBP plot. These stars have extensive emissions in the Ks band because of their wind and mass loss.

\begin{figure}
\centering
\includegraphics[trim=0 0 0 0,clip,width=8.2cm]{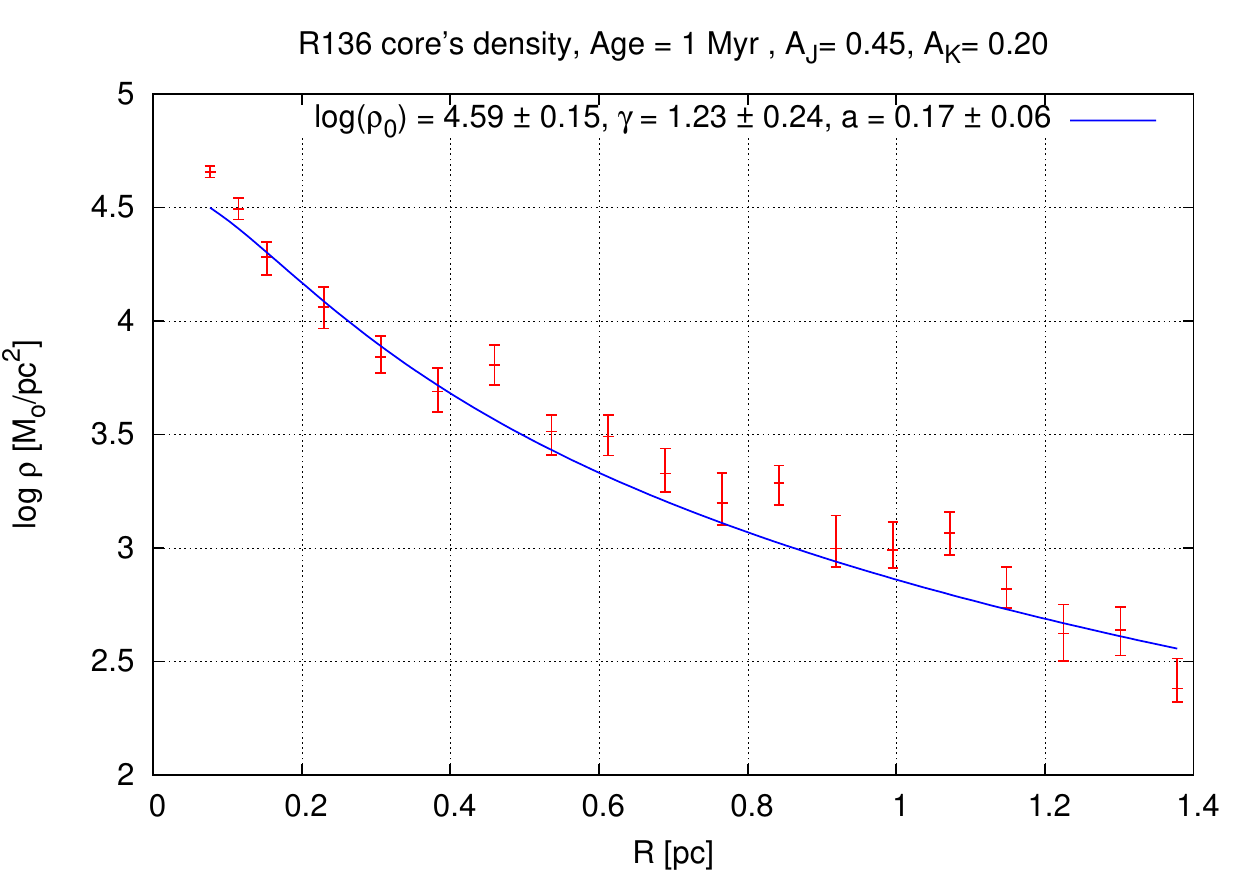}
\caption{Projected mass density [M$_{\odot}$/pc$^2$] profile of R\,136 in IRDIS FoV centered on R136a1. 
The stellar masses are estimated at the age of 1 Myr with extinction values of $A_J=(0.45\pm0.5)~mag$ and $A_K=(0.2\pm0.5)~mag$ in J and Ks band. Eq. \ref{eq:density} is used to fit the blue solid line to the data.
}
\label{fig:density}
\end{figure}

Using the stellar masses estimated at the age of 1 Myr and extinction values in J and K, $A_J=(0.45\pm0.5)~mag$ and $A_K=(0.2\pm0.5)~mag$, we plot the 2D (projected) density profile (Figure \ref{fig:density}). 
We used an Elson-Fall-Freeman (EFF) profile (Elson et al. 1987) to fit the projected mass density in the core of R\,136 (Eq. \ref{eq:density}),

\begin{equation}
\rho [M_{\odot}/pc^2] =  \frac{\rho_0}{(1+\frac{r^2}{a^2})^{\frac{\gamma+1}{2}}}
\label{eq:density}
\end{equation}

We estimated the central mass density of $\rho_0=(3.89^{+1.60}_{-1.14})\times 10^4 [M_{\odot}/pc^2]$ and the parameters $\gamma=1.23\pm0.24$ and $a=0.17\pm0.06$. 
The total observed mass of the clusters for $r < 1.4 pc$ is $M_{obs}=(1.33^{+0.30}_{-0.22})\times10^4 M_{\odot}$, down to 2 $M_{\odot}$. The total mass of the cluster depends on the shape of the MF and the lowest mass limit.
The real stellar masses remain open because we are limited to the angular resolution of SPHERE/IRDIS, so that the estimated central projected density is a lower limit to the real central density. 

We note that the estimated density is projected in 2D. In order to estimate the 3D density approximately, we considered R\,136 to be spherically symmetric and have a radius of $R_{cluster}$. The density profiles were estimated for different $R_{cluster}$ values from 2 pc to 6 pc. In this way, the 3D central densities, $\gamma$, and $a$ were computed by fitting the EFF profile (Eq. \ref{eq:density}).
Table \ref{table:density3d} shows the fitting (3D) parameters for R\,136 considering different values of $R_{cluster}$.
The total mass of the cluster can be estimated by extrapolating to the considered $R_{cluster}$. The ratio of the observed total mass within $r < 1.4~pc$ to the total mass estimated of the cluster within a given radius ($R_{cluster}$) also is given in the last column of Table \ref{table:density3d}.

The estimated value of $\gamma$ and $a$ in 2D and 3D are consistent and the shape of the densities is flatter than the Plummer model (close to the King model). 
All the 3D central densities $\gamma$ and $a$ are lower than the previous values given by Mackey \& Gilmore (2003) and Selman \& Melnick (2013).

\begin{table}
\caption{Estimate of central density of R\,136 in 3D considering different $R_{cluster}$.
The first column gives the hypothetical radius of the cluster. The second column is the 3D central mass density. The third and forth columns are the fitting parameters  $\gamma$ and $a$ in the Eq. \ref{eq:density}. 
Finally, the last column is the ratio of observed mass that is limited by $r < 1.4 ~pc$ to the total mass estimated of the cluster within a given radius. 
$M_{obs}=(1.33^{+0.30}_{-0.22})\times10^4 M_{\odot}$.} 
\centering
\begin{tabular}{c c c c c}  
\small{$R_{cluster}$}&\small{log($\rho_0$)}&$\gamma$&a&\small{$M_{obs}/M_{total}$}\\
$[pc]$&$[M_\odot/pc^3]$&&&\\
\hline
2&$3.86\pm0.16$&$1.24\pm0.23$&$0.18\pm0.06$&$0.91\pm0.02$\\
3&$3.68\pm0.15$&$1.39\pm0.26$&$0.20\pm0.06$&$0.82\pm0.03$\\
4&$3.55\pm0.15$&$1.45\pm0.27$&$0.21\pm0.06$&$0.77\pm0.04$\\
5&$3.46\pm0.16$&$1.49\pm0.28$&$0.22\pm0.07$&$0.74\pm0.04$\\
6&$3.38\pm0.16$&$1.51\pm0.28$&$0.23\pm0.07$&$0.72\pm0.04$\\
\end{tabular} 
\label{table:density3d} 
\end{table}

\section{Discussion and conclusion}
We presented a photometric analysis of the core of R\,136 using the VLT/SPHERE instrument in the near-IR.
The high quality and resolution of these data open a new perspective on our understanding of R136. 
For the first time, more than one thousand sources have been detected in Ks- and J-band data in the small FoV of IRDIS (10.9"$\times$12.3") covering almost 2.7$\times$3.1 pc of R136 core. 
For the ground-based telescopes, the best data come from VLT/MAD, whose the AO quality (Strehl ratios of 15-30\% in Ks) is poorer than that of SPHERE (Strehl ratios of 75\% in Ks). The confusion, especially in the core, therefore remains large enough for the sources that are undetectable in its core.

In SPHERE/IRDIS data, more than 60\% of the stars have companions closer than 0.2" (0.05pc). Ninety percent of the very massive bright stars that have been studied spectroscopically  by Crowther et al. (2016) have visual companions.
The large error bars on the spectroscopic parameters (T$_{eff}$ and logL) prevent us from estimating the age, extinction, and stellar masses accurately.
Using the stellar parameters by Crowther et al. (2016) from the UV spectroscopically analysis, the most probable age of the core is about $1.8^{+1.2}_{-0.8}$ Myr (Figure \ref{fig:histage}), which is consistent with the Crowther et al. (2016) values. 
We find the best age estimate to be about 1 and 1.5 Myr for the very core ($r < 3"$) and outside ($r > 3"$) of R\,136, however, from fitting isochrones with CMD. We clearly see the conjunction of pre-main sequence and main sequence, which is very sensitive to the age.
The extinction in J and Ks are $A_J= (0.45\pm0.5)~mag$ and $A_K=(0.2\pm0.5)~mag$, respectively.
The extinction was estimated from stellar parameters of 54 spectroscopically studied stars by Crowther et al. (2016). The large errors on the extinction values therefore come from the large errors on the spectroscopical analysis on $LogL$ and $T_{eff}$.

Considering the photometric errors, the stellar masses are estimated at different ages with a broad extinction range. 
The MF slope for the 1 and 1.5 Myr isochrone is $\Gamma_{1Myr}=-0.90\pm0.13$ and $\Gamma_{1.5Myr}=-0.98\pm0.18$, respectively, for the mass range of (3 - 300) M$_{\odot}$.

As the core becomes better resolved, more stars are detected. The MF slope is flatter than the Kroupa  ($\Gamma=-1.3$) and Salpeter ($\Gamma=-1.35$) values. Thanks to the SPHERE high-contrast data, we estimated the MF from 3.0 M$_\odot$ up to 300 M$_{\odot}$ at different ages.
The estimated MF slopes are consistent with the values estimated by Malumuth \& Heap (1994), Hunter et al. (1996), and Selman et al. (1999) for the core of R\,136. These values are flatter than the values estimated by Massey \& Hunter (1998) and Brandl et al. (1996). Table \ref{table:slopeshistory} shows the MF slopes estimated for R\,136 in previous works for different mass range and regions. 
The direct comparison is not logical since 1) none of the previous observations, within the IRDIS FoV, covers the mass range of this study and 2) our MF slopes are estimated considering the errors on each stellar mass (see Sect. 4), while the previous studies did not apply the stellar mass errors in the MF.

The derived MF is limited to the resolution of the instrument and also to the detection limit of the observation. Higher angular resolution data may resolve binaries and low-mass stars that affect the shape of the MF. 
Figure \ref{fig:density} shows the density profile of the R136 core at 1 Myr ($0.1pc < r < 1.4 pc$). The lower limit of the central density of R\,136 is $\rho_0=(3.89^{+1.60}_{-1.14})\times 10^4 [M_{\odot}/pc^2]$ at 1 Myr, which is about $\rho_0=288 [stars/pc^2]$. As the mass estimate in the very core of R\,136 is biased by crowding and also for massive WR stars, we excluded the first point from fitting the density profile.
The observed total mass of R\,136 for $r < 1.4 pc$ is $M_{obs}=(1.33^{+0.30}_{-0.22})\times10^4 M_{\odot}$ which covers the mass range of (2 - 300) \msun. This mass is about 51\% of the whole mass range of (0.01 - 300) \msun~ using the Chabrier MF (Chabrier 2003; Chabrier 2005).

Considering that R\,136 is a spherically symmetric cluster with radius $R_{cluster}$ (Table \ref{table:density3d}), we estimated the 3D density profile. The 3D central densities are lower than the values estimated in previous studies given by Mackey \& Gilmore (2003) and Selman \& Melnick (2013). The computed values of $\gamma$ and $a$ (Eq. \ref{eq:density}) are consistent in 2D and 3D considering different $R_{cluster}$. All density profiles are flatter than the Plummer model ($\gamma=4.0$).
 
\begin{figure*}
\centering
\includegraphics[trim=0 0 0 0,clip,width=16cm]{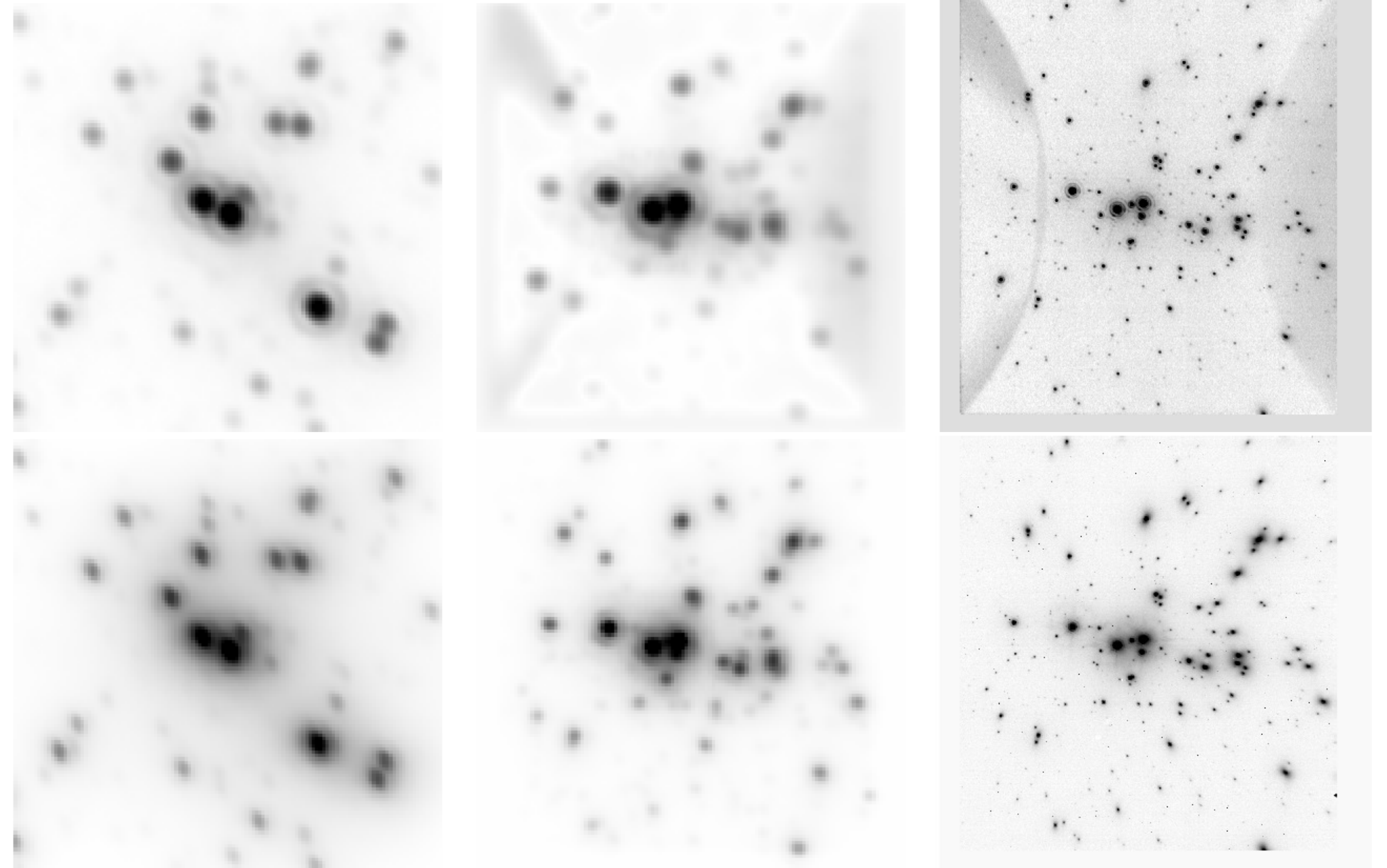}
\caption{Comparison of NGC3603 and R\,136 images from VLT/SPHERE. Left: core of R\,136 (1.56"$\times$1.56") at its real distance. Right: core of NGC3603 (12.5"$\times$12.5") at its real distance. Middle: NGC3603 as it would appear at the same distance as R136. Upper and bottom panels are IRDIS Ks- and J-band images, respectively. 
}
\label{fig:ngc136}
\end{figure*}

\begin{figure*}
\centering
\includegraphics[trim=0 0 0 0,clip,width=8.2cm]{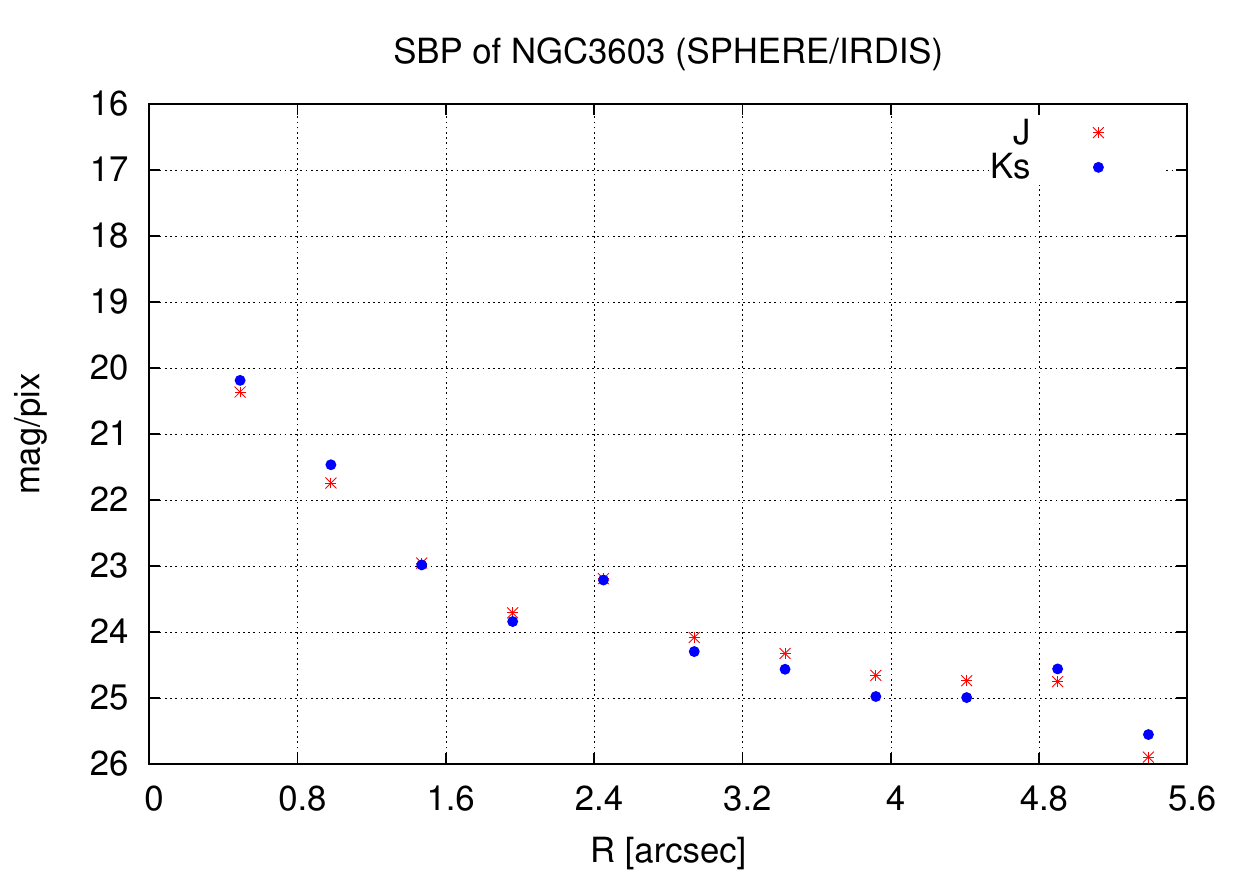}
\includegraphics[trim=0 0 0 0,clip,width=8.2cm]{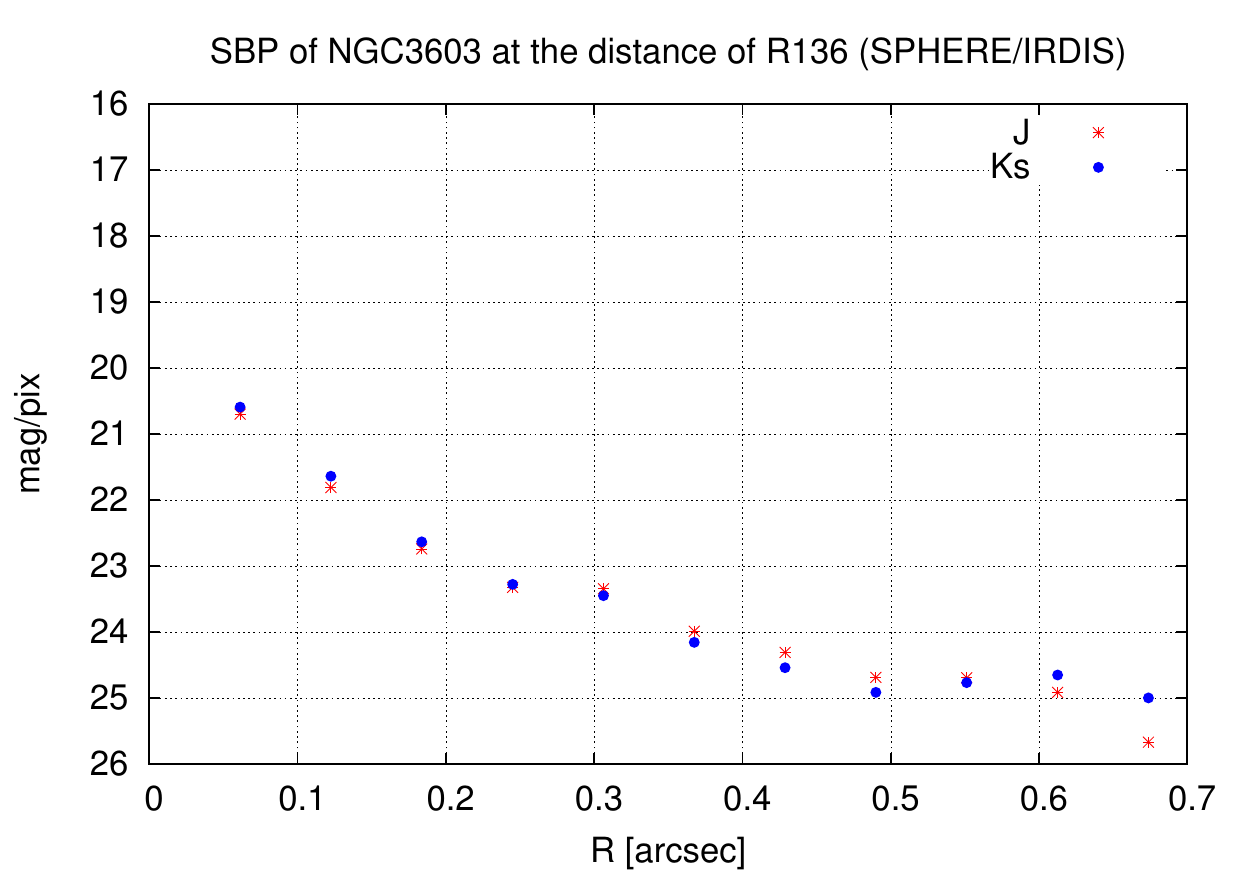}
\caption{SBP of NGC3603 in SPHERE/IRDIS J- and Ks-band data. 
Left: NGC3603 real images. Right: simulated NGC3603 at the distance of R136.
}
\label{fig:sbpngc}
\end{figure*}

Very massive stars in R\,136 have similar characteristics as the galactic WR stars in the core of NGC3603. NGC3603 is almost eight times closer than R136.
The effect of confusion is visible in Figure \ref{fig:ngc136}, in which we visualize NGC3603 at the distance of R\,136 (Figure \ref{fig:ngc136} middle). 
Using {\it{Starfinder}} we detected 408 and 288 stars in J and Ks images, respectively (Khorrami et al. 2016). Using the same criteria for {\it{Starfinder}}, we only detected 109 and 52 sources in J and Ks images of NGC3603 at the distance of R136, which means that more than 70\% of the stars cannot be detected.
This implies that about 1000 detected stars in the R136 core (r < 6") are possibly only 30\% of the real number. The average density in this region (r < 6") would increase from 71 [star/pc$^2$] to 230 [star/pc$^2$]. 
The lack of resolution prevents us from accurately estimating stellar masses, core density, and density profile, while SBP is less affected. 
Figure \ref{fig:sbpngc} shows the SBP of NGC3603 from IRDIS data in J and Ks band, directly and simulated as it would be located at a distance of R136.
The general trend is not affected.

Using SPHERE data, we have gone one step further and partially resolved and understood the core of R\,136, but this is certainly not the final step. 
R\,136 needs to be observed in the future with higher resolution (E-ELT) and/or a more stable PSF (JWST), therefore deeper field imaging. The cluster would then be better characterized for its age, individual and multiple stars, and ultimately, its kinematics on a long enough temporal baseline of observation. 

\begin{table}
	\caption{Mass function slopes for R\,136 from previous analyses.} 
\small
	\begin{tabular}{ l l l } 
		\hline
		MF slope & condition & Reference \\
		\hline
		$-0.90$  & (20 - 70) M$_\odot$ &Malumuth \& Heap 1994\\
		&$ r < 3".3$& \\
		$-1.89$  & (20 - 70) M$_\odot$&Malumuth \& Heap 1994\\
		&$3".3 < r <17".5$ & \\
		$-1.0 \pm 0.1$&(2.8 - 15) M$_\odot$&Hunter et al. 1996 \\
		&$2".0 < r < 18".8$& \\
		$(-1.3) - (-1.4)$&(15 - 120) M$_\odot$&Massey \& Hunter 1998 \\
		$-1.59$ &$r < 1".6  $ &Brandl et al.1996\\
		$-1.33$  &$  1".6<r<3".2 $ &Brandl et al.1996\\
		$-1.63$  &$3".2 < r $ &Brandl et al.1996\\
		$-1.17\pm 0.05 $  &$4".6 < r <19".2$ &Selman et al. 1999 \\
		$-1.37\pm 0.08 $  &$15"<r<75"$ &Selman et al. 1999 \\
		$-1.28 \pm 0.05$& (2 - 6.5) M$_\odot$&Sirianni et al. 2000\\
                                  & $4" \lesssim r \lesssim 20"$&\\
		$-1.2 \pm 0.2$& (1.1 - 20) M$_\odot$ &Andersen et al. 2009\\
				                           & $20"< r < 28"$& \\
		\hline
	\end{tabular} 
	\label{table:slopeshistory} 
\end{table}


{\it{Acknowledgements.}}
{\small{Acknowledgements. ZK is supported by the Erasmus Mundus Joint Doctorate Program by Grant Number 2012-1710 from the EACEA of the European Commission and the University of Nice foundation. The authors are indebted to Dr. A. Stolte for her comments and suggestions for improving the original manuscript. SPHERE is an instrument designed and built by a consortium consisting of IPAG (Grenoble, France), MPIA (Heidelberg, Germany), LAM (Marseille, France), LESIA (Paris, France), Laboratoire Lagrange (Nice, France), INAF– Osservatorio di Padova (Italy), Observatoire de Gen\`eve (Switzerland), ETH Zurich (Switzerland), NOVA (Netherlands), ONERA (France) and ASTRON (Netherlands), in collaboration with ESO. SPHERE was funded by ESO, with additional contributions from CNRS (France), MPIA (Germany), INAF (Italy), FINES (Switzerland) and NOVA (Netherlands). SPHERE also received funding from the European Commission Sixth and Seventh Framework Programmes as part of the Optical Infrared Coordination Network for Astronomy (OPTICON) under grant number RII3-Ct-2004-001566 for FP6 (2004-2008), grant number 226604 for FP7 (2009-2012) and grant number 312430 for FP7 (2013-2016).
		We acknowledge financial support from the Programme National de Plan/'etologie (PNP) and the Programme National de Physique Stellaire (PNPS) of CNRS-INSU. This work has also been supported by a grant from the French Labex OSUG@2020 (Investissements d’avenir – ANR10 LABX56). 
	We thank the anonymous referee for constructive comments, which improved the paper.}}

\bibliographystyle{aa}

\end{document}